\documentclass[usenatbib]{mn2e}
\usepackage{apjfonts,amsfonts,amsmath,amssymb,bm,ctable,verbatim}

\newcommand{\Alf}{{Alfv\'en}}

\newcommand{\bhat}{\hat{\bf b}}

\newcommand{\gizmourl}{\href{http://www.tapir.caltech.edu/~phopkins/Site/GIZMO.html}{\url{http://www.tapir.caltech.edu/~phopkins/Site/GIZMO.html}}}
\newcommand{\FIREurl}{\href{http://fire.northwestern.edu}{\url{http://fire.northwestern.edu}}}

\newcommand{\etal}{et al.}

\newcommand{\acknowledgments}[1]{\begin{small}\section*{Acknowledgments}\end{small}{\noindent #1}\vspace{5pt}}
\newcommand{\datastatement}[1]{\begin{small}\section*{Data Availability Statement}\end{small}{\noindent #1}\vspace{5pt}}

\title[Sub-Grid CR Transport]{A Simple Sub-Grid Model For Cosmic Ray Effects on Galactic Scales}

\author[Hopkins \etal]{
\parbox[t]{\textwidth}{
Philip F.~Hopkins$^1$, Iryna S.\ Butsky$^{1}$, Suoqing Ji$^{2,1}$, Du\v{s}an Kere\v{s}$^{3}$
}\vspace*{4pt} \\
$^1$ TAPIR, Mailcode 350-17, California Institute of Technology, Pasadena, CA 91125, USA. E-mail:phopkins@caltech.edu \\
$^2$ Astrophysics Division \&\ Key Laboratory for Research in Galaxies and Cosmology, Shanghai Astronomical Observatory, \\
\ \ \ \ \ \ \ \ Chinese Academy of Sciences, 80 Nandan Road, Shanghai 200030, China \\
$^3$ {Department of Physics, Center for Astrophysics and Space Science, University of California at San Diego, 9500 Gilman Drive, La Jolla, CA 92093, USA}
}

\date{}
\begin{document}
\maketitle

\begin{abstract}
Many recent numerical studies have argued that cosmic rays (CRs) from supernovae (SNe) or active galactic nuclei (AGN) could play a crucial role in galaxy formation, in particular by establishing a CR-pressure dominated circum-galactic medium (CGM). But explicit CR-magneto-hydrodynamics (CR-MHD) remains computationally expensive, and it is not clear whether those results can be applied to simulations that do not explicitly treat magnetic fields or resolved ISM phase structure. We therefore present an intentionally extremely-simplified ``sub-grid'' model for CRs, which attempts to capture the key qualitative behaviors of greatest interest for those interested in simulations or semi-analytic models including some approximate CR effects on galactic ($\gtrsim$\,kpc) scales, while imposing negligible computational overhead. The model is numerically akin to some recently-developed sub-grid models for radiative feedback, and allows for a simple constant parameterization of the CR diffusivity and/or streaming speed; it allows for an arbitrary distribution of sources (proportional to black hole accretion rates or star-particle SNe rates or gas/galaxy star formation rates), and interpolates between the limits where CRs escape the galaxies with negligible losses and those where CRs lose most of their energy catastrophically before escape (relevant in e.g.\ starburst galaxies). The numerical equations are solved trivially alongside gravity in most codes. We compare this to explicit CR-MHD simulations and discuss where the (many) sub-grid approximations break down, and what drives the major sources of uncertainty.
\end{abstract}


\section{Introduction}
\label{sec:intro}

In recent years, a number of different studies have shown by explicitly evolving CR dynamics coupled to the MHD equations in galaxy-formation simulations, that CRs could play a key role in galaxy formation \citep{jubelgas:2008.cosmic.ray.outflows,uhlig:2012.cosmic.ray.streaming.winds,Wien13,salem:2013.cosmic.ray.outflows,Simp16,Pakm16,Rusz17,Giri18}. Most notably, CRs from SNe or AGN could provide an additional source of pressure in the halo or circum-galactic medium (CGM) which can suppress new inflows of cooling gas or re-accelerate outflows to inter-galactic medium (IGM) scales, significantly altering galaxy formation \citep{Sale16,Buts18,butsky:2020.cr.fx.thermal.instab.cgm,su:turb.crs.quench,hopkins:cr.mhd.fire2,ji:fire.cr.cgm,holguin:2019.cr.streaming.turb.damping.cr.galactic.winds,su:2018.stellar.fb.fails.to.solve.cooling.flow,su:turb.crs.quench,hopkins:cr.mhd.fire2,buck:2020.cosmic.ray.low.coeff.high.Egamma,hopkins:2020.cr.outflows.to.mpc.scales,hopkins:cr.transport.constraints.from.galaxies,hopkins:2020.cr.transport.model.fx.galform,ji:fire.cr.cgm,ji:20.virial.shocks.suppressed.cr.dominated.halos}.

But it is computationally very expensive to explicitly incorporate CR transport in numerical simulations, and adds substantial computational complexity. Like radiation-hydrodynamics (RHD), CRs represent a broad spectral distribution (so one would ideally desire to evolve a range of CRs momenta or rigidities, like wavelengths of light, as in \citealt{girichidis:cr.spectral.scheme,ogrodnik:2021.spectral.cr.electron.code,hopkins:cr.multibin.mw.comparison,hopkins:2021.sc.et.models.incompatible.obs,hopkins:cr.spectra.accurate.integration}), and move locally at up to the speed of light. One can reduce the complexity by integrating over energies to obtain a ``single-bin'' approximation (for spectrally-integrated quantities like the total CR energy or pressure), and using a moments-based method obtain an energy and flux (``M1-like'') or pure energy (``FLD-like'' or Fokker-Planck) equation for the CRs \citep{Zwei13,chan:2018.cosmicray.fire.gammaray,jiang.oh:2018.cr.transport.m1.scheme,thomas.pfrommer.18:alfven.reg.cr.transport,hopkins:m1.cr.closure}. However, solving such equation explicitly still imposes severe timestep costs, e.g.\ explicitly integrating CR diffusion requires a timestep $\Delta t \le C\,\Delta x^{2}/\kappa$ where $\kappa$ is the diffusivity, $\Delta x$ the numerical resolution, and $C$ a Courant-like factor. For the observationally-required values of $\kappa$ (at the energies $\sim 1-10\,$GeV which dominate the CR pressure) needed to reproduce Solar system CR observations \citep{blasi:cr.propagation.constraints,vladimirov:cr.highegy.diff,gaggero:2015.cr.diffusion.coefficient,2016ApJ...819...54G,2016ApJ...824...16J,cummings:2016.voyager.1.cr.spectra,2016PhRvD..94l3019K,evoli:dragon2.cr.prop,2018AdSpR..62.2731A,hopkins:cr.multibin.mw.comparison,delaTorre:2021.dragon2.methods.new.model.comparison} and extragalactic $\gamma$-ray constraints \citep{chan:2018.cosmicray.fire.gammaray,su:turb.crs.quench,hopkins:cr.transport.constraints.from.galaxies,hopkins:cr.mhd.fire2,hopkins:2020.cr.transport.model.fx.galform,bustard:2020.crs.multiphase.ism.accel.confinement}, this translates to $\Delta t \lesssim (\Delta x/{\rm pc})^{2}\,{\rm yr}$. Higher-moment methods can remove this constraint at the expense of introducing the speed of light as a signal speed ($\Delta t \le C\,\Delta x/c$), then employing a ``reduced speed of light'' (RSOL) approximation, but this still requires the RSOL be much faster than any other signal speeds to obtain converged solutions \citep{chan:2018.cosmicray.fire.gammaray}, which by definition significantly reduces the timesteps. And while implicit numerical methods \citep[e.g.][]{sharma.2010:cosmic.ray.streaming.timestepping,sharma.2011:fast.semi.implicit.anisotropic.diffusion,kannan.2015:anisotropic.conduction.arepo,Pakm16} can be stably integrated for somewhat longer timesteps they impose their own (often very large) overhead costs and typically scale poorly if a deep hierarchical timestep structure is adopted \citep{hopkins:gizmo.diffusion}. 

Moreover, it is not clear whether the cost or complexity of adding explicit CR transport is worthwhile, in simulations which neglect other key physics which are important for CR transport, losses, and coupling to gas, such as magnetic field dynamics, variation in local radiation energy densities, explicitly-resolved neutral vs.\ ionized phases of the ISM, and individually-time-resolved SNe and/or collimated AGN jets. In simulations like those in e.g.\ \citet{vogelsberger:2013.illustris.model,crain:eagle.sims,2017MNRAS.467..179G,pillepich:2018.illustris.tng,buck:2020.cosmic.ray.low.coeff.high.Egamma}, it is common practice to treat other forms of stellar and AGN feedback via sub-grid prescriptions that treat the ISM with some ``effective equation of state'' instead of resolving its structure, and insert key effects of SNe and AGN mechanical feedback ``by hand'' (e.g.\ injecting thermal energy, ``kicking'' particles into outflows, or turning off cooling for some specified period of time). Ideally, these models for mechanical feedback are calibrated directly to the results of higher-resolution simulations that attempt to actually resolve those phenomena. Clearly, one would like to have a similar treatment for CRs. However, the popular sub-grid model approaches above cannot appropriately treat CRs: if the dominant effect of CRs is the introduction of non-thermal pressure terms, and their gradients are important on large scales like the CGM (far from their injection sites), then one cannot qualitatively approximate this with some ``thermal+kinetic'' or ``cooling turnoff'' type model. 

In this paper, therefore, we attempt to develop a simple sub-grid prescription for use in simulations of galactic or cosmological scales (specifically, simulations which do not attempt to resolve ISM phases or explicitly treat CR transport). Our goal is to design the {\em simplest possible} toy model which can capture the most important qualitative effects/behaviors of CRs insofar as they influence galaxy formation and CGM/IGM structure, which can be incorporated into simulations like those mentioned above with essentially zero computational cost, and allow users to parameterize the CR transport parameters (whose detailed scaling is probably the most uncertain parameter governing CR effects on galaxies, see  \citealt{Buts18,butsky:2020.cr.fx.thermal.instab.cgm,hopkins:2020.cr.transport.model.fx.galform}) in a simple manner. We stress that this is in no way a replacement for simulations which do explicitly model these physics: such simulations are necessary and crucial to inform models like those here, as well as to actually make quantitative predictions for CR observables (which the model we propose below is {\em not} appropriate for) in order to actually constrain the detailed role of CRs in galaxy formation.

\section{Derivation}
\label{sec:deriv}

\subsection{Local CR Energy Density \&\ Pressure}
\label{sec:deriv:energy}

\subsubsection{Generalized Large-Scale CR Transport Equations}

Beginning from the fully-general CR transport Vlasov equation, one can make a series of assumptions and transformations to progressively make the equations simpler. First, assume Lorentz forces rapidly ensure a microscopically near-gyrotropic CR distribution function $f$, to obtain the focused CR transport equation \citep[][]{2015ApJ...801..112L} which is valid to leading order in $\mathcal{O}(r_{g}/L)$ (where $r_{g}$ is the CR gyro-radius, and $L$ the macroscopic resolved scales in the simulations), with the standard quasi-linear theory scattering rate coefficients from \citet{schlickeiser:89.cr.transport.scattering.eqns}. 

Next, take the first and second pitch-angle moments equations with an interpolated closure relation and expand to leading order in $\mathcal{O}(u/c)$ where $u$ is the (non-relativistic) background fluid velocity to obtain the moments equations in \citet{hopkins:m1.cr.closure}. Then integrate these over momentum-space, imposing the assumption of a fixed spectral shape for the CRs with most of the energy in ultra-relativistic CRs, to obtain the energy equations, Eq.~38 in \citealt{hopkins:m1.cr.closure}:
\begin{align}
\nonumber D_{t} e_{\rm cr}  &+ \nabla \cdot ( F_{e}\,\bhat ) \approx S_{e} - \mathbb{P}_{\rm cr}:\nabla{\bf u}
-\frac{\bar{\nu}}{c^{2}}\left[ \bar{v}_{A}\,F_{e} - 3\,\chi\,{v}^{2}_{A}\,\left( e_{\rm cr} + P_{\rm cr} \right)  \right], \\
\label{eqn:e.total}  D_{t} F_{e}  &+ c^{2}\, \bhat \cdot ( \nabla \cdot \mathbb{P}_{\rm cr} ) \approx 
-\bar{\nu}\,\left[ F_{e} - 3\,\chi\,\bar{v}_{A}\,(e_{\rm cr}+P_{\rm cr}) \right] + S_{F_{e}}\ ,
\end{align}
where $e_{\rm cr}$, $P_{\rm cr} \approx e_{\rm cr}/3$, $\mathbb{P}_{\rm cr}$, $F_{e}$ are the CR energy density, scalar pressure, pressure tensor, and energy flux; ${\bf u}$ is the gas fluid velocity; $D_{t}X \equiv \partial_{t} X + \nabla \cdot ({\bf u}\,X)$ is the conservative comoving derivative; $S_{e}$ and $S_{F_{e}}$ represent sources and sinks; $\bhat$ is the unit magnetic field vector $={\bf B}/|{\bf B}|$; $\bar{\nu}$ is the appropriately spectrally-averaged mean CR scattering rate ($=\bar{\nu}_{+}+\bar{\nu}_{-}$, the sum of contributions from forward and backward-propagating waves); $c$ is the speed of light; $v_{A}$ is the appropriate \Alf\ speed (\Alf\ speed of modes with wavenumber $\sim 1/r_{g}$ where $r_{g}$ is the gyro radius); $\bar{v}_{A} \equiv v_{A}\,(\bar{\nu}_{+} - \bar{\nu}_{-})/(\bar{\nu}_{+}+\bar{\nu}_{-})$ is the signed ``streaming speed,'' and $\chi\equiv (1-\langle \mu^{2}\rangle)/2$ is a completely general closure function that defines the (an)isotropy of $\mathbb{P}_{\rm cr}\equiv 3\,P_{\rm cr}\,[\chi\,\mathbb{I} + (1-3\chi)\,\bhat\bhat ]$ for any gyrotropic CR $f$ in terms of the second moment of the pitch-angle $\langle \mu^{2} \rangle$.

Next, take this and assume the CRs have reached flux-steady-state in the strong-scattering limit, i.e.\ $D_{t} F_{e} \rightarrow 0$ (or equivalently take the Newtonian limit, $c\rightarrow \infty$), which occurs in a scattering time $\sim \bar{\nu}^{-1} \sim 30\,$yr for $\sim 1\,$GV CRs \citep{hopkins:cr.transport.constraints.from.galaxies}, leading to the strong-scattering limit with a close-to-isotropic CR distribution function ($\chi\rightarrow 1/3$, $\mathbb{P}_{\rm cr} \rightarrow P_{\rm cr}\,\mathbb{I}$). In this limit $\bar{v}_{A}$ approaches one of two limits: either $\bar{v}_{A} \rightarrow 0$ if CR scattering is symmetric in the \Alf\ frame, or $\bar{v}_{A} \rightarrow -{\rm SIGN}(\bhat\cdot \nabla P_{\rm cr})\, v_{A}$ if the scattering is asymmetric (as expected if modes excited by the CRs dominate scattering as in self-confinement models). This gives a CR energy equation we can write as:
\begin{align}
\partial_{t} e_{\rm cr} \rightarrow & \nabla \cdot \left[ \frac{c^{2}}{3\,\bar{\nu}}\,\bhat\bhat \nabla e_{\rm cr} - (\bar{v}_{A}\,\bhat + {\bf u})\,e_{\rm cr} \right]  \\
\nonumber & - P_{\rm cr} \nabla\cdot(\bar{v}_{A}\,\bhat + {\bf u}) + \frac{v_{A}^{2}-\bar{v}_{A}^{2}}{c^{2}}\,\bar{\nu}\,(e_{\rm cr} + P_{\rm cr}) + S_{e}\ .
\end{align}

\subsubsection{Isotropized Steady-State Equations}

Now we make a series of much stronger assumptions akin to those used to derive the commonly-adopted (see e.g.\ \citealt{strong:2001.galprop,evoli:dragon2.cr.prop}) steady-state isotropic Fokker-Planck equation for CRs: (1) assume the energy equation is in Eulerian steady-state ($\partial_{t} e_{\rm cr} \rightarrow 0$, which occurs on a bulk CR transport/injection timescale $\sim L^{2}/\kappa$ or $e_{\rm cr}/S$, which can range from $\sim 10^{7}\,$yr in the dense ISM to $\sim 1-10\,$Gyr in the CGM), and (2) that the magnetic fields are isotropically ``tangled'' on scales of order the CR scattering mean-free-path and below the resolution scale, and we implicitly replace all quantities with their averages over the resolution-scale ``tangling,'' so that we can replace the anisotropic diffusion tensor $\boldsymbol{\kappa}_{\|} \equiv \kappa_{\|}\,\bhat\bhat = (c^{2}/3\,\bar{\nu})\,\bhat\bhat$ with an isotropic equivalent\footnote{For a more formal justification of this, see e.g.\ \citet{braginskii:viscosity,berezinskii:cr.book,Zwei13}, and for a more practical example showing that this is an acceptable approximation at the order-of-magnitude level in simulations using explicitly anisotropic transport (including those studied here below), see \citet{chan:2018.cosmicray.fire.gammaray,su:2018.stellar.fb.fails.to.solve.cooling.flow,su:turb.crs.quench,hopkins:cr.mhd.fire2,ji:fire.cr.cgm,buck:2020.cosmic.ray.low.coeff.high.Egamma}.} $\langle \nabla \cdot (\boldsymbol{\kappa}_{\|} \cdot \nabla e_{\rm cr}) \rangle \sim \nabla (\kappa_{\rm iso}\,\nabla e_{\rm cr})$ with $\kappa_{\rm iso} \sim \kappa_{\|}/3 = c^{2}/9\,\bar{\nu}$, and likewise replace ${\bf v}_{\rm st} \rightarrow {\bf u} + (|\bar{v}_{A}|/\sqrt{3})\,\hat{\nabla} P_{\rm cr}$. With these, and replacing $P_{\rm cr} \approx e_{\rm cr}/3$, we obtain
\begin{align}
-\nabla \cdot \left[ \kappa_{\rm iso} \nabla e_{\rm cr} - {\bf v}_{\rm st}\,e_{\rm cr} \right] 
&\sim -\frac{\nabla \cdot {\bf v}_{\rm st}}{3}\,e_{\rm cr} + \frac{4\,(v_{A}^{2}-\bar{v}_{A}^{2})}{27\,\kappa_{\rm iso}}\,e_{\rm cr} + S\ .
\end{align}
Here the terms on the left-hand side represent the ``diffusive'' transport (in $\kappa$) and the ``streaming'' plus ``advective/convective'' transport (in ${\bf v}_{\rm st}$) -- it is clear from the equation above (though see Appendix~B in \citealt{hopkins:cr.transport.constraints.from.galaxies} for more detailed discussion) that these are only meaningfully separable into classical ``diffusion'' versus ``streaming'' behaviors if $\kappa_{\rm iso}$ and ${\bf v}_{\rm st}$ are strictly constants: if we allow them to be arbitrary functions of position, then the coefficients are strictly degenerate (as we note below) and one cannot mathematically separate the two terms. On the right-hand side, we have the ``adiabatic'' and ``streaming loss'' term (in $\nabla \cdot {\bf v}_{\rm st}$), the ``diffusive reacceleration'' term (in $(v_{A}^{2}-\bar{v}_{A}^{2})$) and the $S$ collects sources (at injection sites) and losses. Note that we define ${\bf v}_{\rm st}$ to collect the ``streaming'' and advective/convective/adiabatic terms together on both sides (different from the usual convention), because our model is ultimately defined in an Eulerian frame of the galaxy.

\subsubsection{Spherically-Symmetric Form Away From a Point Source}

Next, separate $S \rightarrow j_{\rm in} - Q_{\rm loss}$ with injection $j_{\rm in}$ and losses $Q_{\rm loss} \approx \Psi_{\rm loss}\,e_{\rm cr}$ for some loss rate function $\Psi_{\rm loss}$,\footnote{For CR spectra dominated by $\sim1-10\,$GV protons with observationally-favored transport speeds (much faster than \Alf{ic}), the CR-spectrum-integrated loss function $\Psi_{\rm loss}$ is dominated by a combination of hadronic/pionic, Coulomb, and ionization losses, scaling as 
\begin{align}
\Psi_{\rm loss} &\approx 10^{-16}\,{\rm cm^{3}\,s^{-1}}\,(6.4\,n_{n} + 3.1\,n_{e} + 1.8\,n_{\rm HI})
\end{align}
where $n_{n}$, $n_{e}$, and $n_{\rm HI}$ are the number densities of nucleons, free electrons, and neutral atoms, respectively \citep{Mann94}.}
 and consider a point source with location defined as the origin ($j_{\rm in} = \dot{E}_{\rm cr}\,\delta({\bf x})$). Now make one more series of strong assumptions: take $v_{\rm st} \equiv |{\bf v}_{\rm st}|$ to be approximately constant, in a spherically-symmetric ambient medium (so $\Psi \rightarrow \Psi(r,\,...)$), giving (away from $r=0$ aka outside of sources, so $j_{\rm in} \rightarrow 0$):
\begin{align}
\label{eqn:ecr.spherical} \frac{1}{r^{2}}\,\frac{\partial }{ \partial r} \left\{ r^{2}\,\left[ v_{\rm st}\,e_{\rm cr} - \kappa_{\rm iso}\,\frac{\partial e_{\rm cr}}{\partial r} \right] \right\}
&  =  \\
\nonumber - \frac{2\,v_{\rm st}\,e_{\rm cr}}{3\,r} &+  \left[\frac{4\,(v_{A}^{2} - \bar{v}_{A}^{2})}{27\,\kappa_{\rm iso}} - \Psi_{\rm loss}  \right]\,e_{\rm cr} \ .
\end{align}

\subsubsection{The ``Streaming+Diffusion'' Approximation with Constant Coefficients}

Even with all our simplifying assumptions and spherical symmetry, with $\langle \kappa_{\rm iso} \rangle = \langle \kappa_{\rm iso} \rangle(r)$, Eq.~\ref{eqn:ecr.spherical} can only be solved numerically making a variety of additional assumptions about the form of $\kappa_{\rm iso}$ and $\bar{v}_{A}$, $v_{A}$, etc. However, it is useful at this point to note that for self-confinement models, the ``diffusive reacceleration'' term in $v_{A}^{2}-\bar{v}_{A}^{2}$ vanishes identically, and even for extrinsic turbulence models (where $\bar{v}_{A} \rightarrow 0$ so this re-acceleration term is maximized), the term is orders-of-magnitude smaller than the $\Psi_{\rm loss}$ term for realistic values of $\kappa_{\rm iso}$ (see discussion in \citealt{hopkins:m1.cr.closure,hopkins:cr.multibin.mw.comparison}). We can therefore drop it safely.\footnote{We stress that there is no inherent conflict between dropping this term here, and claims in some steady-state analytic Galactic cosmic ray transport models (with e.g.\ GALPROP, see \citealt{korsmeier:2022.cr.fitting.update.ams02}) that there is a ``preference'' for including diffusive reacceleration (DRA). Taking, for example, the favored parameters from the model variant with the strongest DRA in \citet{korsmeier:2022.cr.fitting.update.ams02}, then the DRA term ($\propto v_{A}^{2} - \bar{v}_{A}^{2}$) in Eq.~\ref{eqn:ecr.spherical} is approximately $\sim 10^{5}$ times smaller than the leading-order term in Eq.~\ref{eqn:ecr.spherical}. Importantly, (1) there are many small correction terms which might manifest in very detailed models of full CR spectra of many species, which do not dominate the leading-order uncertainties in the transport of most of the CR energy to the CGM (the only quantity our models really attempt to capture); (2) the leading-order effect of DRA on the CR spectra is diffusion in momentum-space (the usual ``$D_{pp}$'' term) which can alter CR spectra shapes but has a much weaker effect on the bulk transport of total CR energy; and (3) such terms are at least partially degenerate with terms that cannot be captured in steady-state models, such as the adiabatic term, which most CR-MHD simulations find to be significantly larger, as we discuss below \citep{2017ApJ...847L..13P,chan:2018.cosmicray.fire.gammaray,buck:2020.cosmic.ray.low.coeff.high.Egamma,butsky:2020.cr.fx.thermal.instab.cgm,hopkins:cr.multibin.mw.comparison}.}
 Next note that the $v_{\rm st}$ term includes only the ``advective'' and \Alf\ velocities: $v_{\rm st} = |{\bf u} + \bar{v}_{A}\,\bhat/\sqrt{3}|$, where $|\bar{v}_{A} | < v_{A}$, which is much smaller than the effective ``bulk'' streaming or diffusion or transport speed of CRs (given by $v_{\rm eff} \sim |\kappa_{\rm iso}\,e_{\rm cr}^{-1}\,\nabla e_{\rm cr}|$) at least {\em within} the galaxy for any observationally-allowed diffusivities in Milky Way like and dwarf galaxies at the energies ($\sim 1-10\,$GeV) which dominate the total CR energy density $e_{\rm cr}$ \citep[see][]{blasi:cr.propagation.constraints,vladimirov:cr.highegy.diff,gaggero:2015.cr.diffusion.coefficient,2016ApJ...819...54G,2016ApJ...824...16J,cummings:2016.voyager.1.cr.spectra,2016PhRvD..94l3019K,evoli:dragon2.cr.prop,2018AdSpR..62.2731A,chan:2018.cosmicray.fire.gammaray,su:turb.crs.quench,hopkins:cr.transport.constraints.from.galaxies,hopkins:cr.mhd.fire2,hopkins:2020.cr.transport.model.fx.galform,bustard:2020.crs.multiphase.ism.accel.confinement,delaTorre:2021.dragon2.methods.new.model.comparison}, though we will allow for some streaming outside the galaxy as discussed below. Typically in these studies the inferred $v_{\rm eff} \sim (10^{2} - 10^{4})\,v_{A}$. This means we can neglect the ``streaming+adiabatic loss term'' $2\,v_{\rm st}\,e_{\rm cr}/(3\,r) \lesssim v_{A}/r$ as a dominant loss term for $e_{\rm cr}$ (though we note below the ``adiabatic'' part of this term can sometimes be non-negligible).\footnote{This statement regarding $v_{\rm eff}$ is somewhat radius ($r$)-dependent: for constant streaming/diffusion coefficients, the streaming term will be less important at small $r$ and more important at large-$r$. We discuss how large a correction this can be in more detail in \S~\ref{sec:assumptions} below.} 

However, it is common in the literature to refer to ``super-\Alf{ic} streaming'' arising from self-confinement motivated CR transport models. In these models, the scattering rate $\nu$ (and therefore $\kappa$) depends itself on $e_{\rm cr}$ and its gradients, so it introduces ``streaming like'' behavior into $\kappa_{\rm iso}$. To capture this to leading order, in our spherically-symmetric, time-invariant steady-state approximation, we assume we can expand the spherically-averaged, direction-averaged effective diffusivity as $\langle \kappa_{\rm iso} \rangle \sim \langle c^{2}/9\,\bar{\nu}(r,\,...) \rangle \sim \kappa_{0} + v_{\kappa}\,r/2 + ...$, where $\kappa_{0}$ and $v_{\kappa}$ are constants (and we write $v_{\kappa} \equiv 2\,(\partial \langle \kappa_{\rm iso}[r] \rangle / \partial r)$ because this term has units of velocity). This separates the behaviors of the scattering term $\kappa_{\rm iso}$ into an effectively traditional ``diffusion-like'' term $\kappa_{0}$ and a ``streaming-like'' term $v_{\kappa}$. 

We stress that as shown in \citet{hopkins:cr.transport.constraints.from.galaxies}, this is only an approximation to the mean behavior of CR scattering models: actual self-confinement models, for example, produce coefficients which are complicated functions of $e_{\rm cr}$ and its derivatives in a manner which means that, in detail, the behavior of Eq.~\ref{eqn:ecr.spherical} is neither truly that of a ``diffusion'' or a ``streaming'' equation. But large uncertainties remain in these models, so we choose to simply parameterize the coefficients as above. With this, we define the ``effective streaming speed'' $v_{\rm st,\,eff} \equiv v_{\kappa} + \bar{v}_{A} \approx v_{\kappa}$. Note that here and throughout this paper, we use ``streaming'' to refer to any transport term which produces ``streaming-like dynamics'' in this sense, regardless of whether it originates via large-scale gradients in the scattering rate or $\kappa$ with position, or \Alf{ic} streaming, or free-streaming at $c$ or other processes. In the models here, these are all degenerate (they all produce essentially identical predictions), so we are agnostic to their physical origins. However we emphasize this is a different usage of the term ``streaming'' than in some of the historical cosmic ray literature.

\subsubsection{Approximate Closed-Form Analytic Expression}

Now with all of these approximations, (which reduce Eq.~\ref{eqn:ecr.spherical} to $r^{-2}\,\partial_{r}\{ r^{2}\,[ v_{\rm st}\,e_{\rm cr} - \kappa\,\partial_{r} e_{\rm cr}] \} = -\Psi\,e_{\rm cr}$), it becomes possible to solve Eq.~\ref{eqn:ecr.spherical} numerically, but it still does not have closed-form analytic solutions. However given the gross approximations we have already made, we can approximate the exact numerical solutions to more than sufficient accuracy (and capture all of the relevant limiting behaviors within the context allowed by our approximations) with the following simple closed-form solution:\footnote{Eq.~\ref{eqn:ecr.single.source} comes from combining the reduced Eq.~\ref{eqn:ecr.spherical} with $v_{\rm st,\,eff} \equiv v_{\rm st,\,0} + v_{\kappa}$ and $\kappa \equiv \kappa_{0} + v_{\kappa}\,r/2$, together with the flux boundary condition in spherical symmetry $\oint {\bf F}_{\rm cr}\,\cdot d{\bf A} {|}_{r\rightarrow 0} = 4\pi\,r^{2}\, (v_{\rm st,\,0}\,e_{\rm cr} - \kappa\,\partial_{r} e_{\rm cr}) |_{r\rightarrow 0} = \dot{E}_{\rm cr}$. There is no closed-form exact analytic solution, but it is easy to verify by insertion that Eq.~\ref{eqn:ecr.single.source} satisfies both constraint equations in each of the four possible limits: (1) diffusion-dominated ($\kappa_{0} \gg v_{\rm st,\,eff}\,r$ or $v_{\rm st,\,eff} \rightarrow 0$) with weak losses ($\Psi\rightarrow 0$); (2) diffusion dominated with strong losses ($\Psi$ large, so $\partial_{r}\,e_{\rm cr} \approx -\psi\,e_{\rm cr}$ to leading order); (3) ``streaming'' dominated ($\kappa_{0} \ll v_{\rm st,\,eff}\,r$ or $\kappa_{0} \rightarrow 0$) with weak losses; (4) ``streaming'' dominated with strong losses. And it is trivial to verify from the form of Eq.~\ref{eqn:ecr.single.source} that the transition between each of these regimes occurs at the order-of-magnitude value where we would expect (i.e.\ between diffusion and streaming dominated when $v_{\rm st,\,eff}\,r \sim \kappa_{0}$, or between negligible losses and loss-dominated when $\int \psi\,dr \sim 1$, equivalent to the statement that the transport and loss times out to some radius $r$ are roughly equal.}
\begin{align}
\label{eqn:ecr.single.source} e_{\rm cr} &\approx \frac{\langle \dot{E}_{\rm cr} \rangle_{t_{\rm r}}}{4\pi\,r\,(\kappa_{0} + {v}_{\rm st,\,eff}\,r)}\,\exp{\left\{ -\int_{0}^{r}\,\psi_{\rm loss}\,dr \right\}} \\
\nonumber &=  \frac{\langle\dot{E}_{\rm cr} \rangle_{t_{\rm r}}\,e^{-\tau_{\rm cr}}}{4\pi\,r\,(\kappa_{0} + {v}_{\rm st,\,eff}\,r)} \\ 
\label{eqn:psi.loss} \psi_{\rm loss} &\approx \frac{\Psi_{\rm loss}}{v_{\rm st}}\,\left(1 + \frac{\kappa\,\Psi_{\rm loss}}{v_{\rm st}^{2}} \right)^{-1/2} 
\end{align}
where we use $\langle \dot{E}_{\rm cr} \rangle_{t_{\rm r}}$ to denote the time-averaged value of the source function $\dot{E}_{\rm cr}$ averaged over approximately the bulk CR transport time (since the CR energy cannot instantly adjust to small-timescale variations in $\dot{E}_{\rm cr}$). 

We can then approximate $e_{\rm cr}$ by integrating over all sources. 

\subsection{Couplings to Gas}
\label{sec:deriv:coupling}

If we again begin from the general CR-gas coupling terms described in \citet{hopkins:cr.multibin.mw.comparison,hopkins:m1.cr.closure}, including Lorentz forces, scattering, ionization and Coulomb and catastrophic interactions, etc., and take all of the same limits assumed above in \S~\ref{sec:deriv:energy}, then the remaining leading-order coupling terms to the gas momentum and energy equations can be written entirely in terms of $e_{\rm cr}$ derived above. 

In the gas momentum equation, for the limits above, the CR pressure tensor $\mathbb{P}_{\rm cr} \approx P_{\rm cr}\,\mathbb{I} \approx (e_{\rm cr}/3)\,\mathbb{I}$ is approximately isotropic and simply adds to the total isotropic gas pressure in the momentum equation as $P_{\rm gas} \rightarrow P_{\rm gas,\,non-cr} + P_{\rm cr}$. 

In the gas thermo-chemistry, the CRs give rise to an ionization rate (for our assumed universal CR spectral shape) expressed in terms of the usual $\zeta$ parameter as $\zeta_{\rm cr} \approx 5\times 10^{-18}\,{\rm s^{-1}}\,(e_{\rm cr}/{\rm eV\,cm^{-3}})$ (note in some conventions $\zeta_{\rm cr}$ must be multiplied by $\sim 1.5$ for atomic gas and $\sim 2.3$ for molecular gas). Though we caution that the ionization rate depends primarily on low-energy ($\sim$\,MeV) CRs, which contribute little to the total energy, so there could be large variations in this owing to un-modeled variations in the CR spectrum (and there appears to be direct evidence for this in Milky Way GMCs; \citealt{indriolo:2015.cr.ionization.rate.vs.galactic.radius}). Still, this can provide a substantial improvement on the common practice of simply assuming a single uniform-in-space-and-time $\zeta_{\rm cr}$.

In the gas thermal energy equation, in addition to ionization heating parameterized via $\zeta$ above, the energy lost by CRs in hadronic/pionic and Coulomb interactions is partially thermalized (and partially lost to escaping radiation/particles such as $\gamma$-rays; see  \citealt{Mann94,guo.oh:cosmic.rays}), giving $\dot{e}_{\rm th,\,gas} \approx e_{\rm cr}\,(0.9\,n_{n} + 1.6\,n_{e})\times10^{-16}\,{\rm s^{-1}}$ (where $n_{n}$ is the nucleon number density and $n_{e}$ the free-electron number density). Optionally, if one assumes that the CR scattering is dominated by CR self-excited waves such that $|\bar{\nu}_{+} - \bar{\nu}_{-}|/|\bar{\nu}_{+}+\bar{\nu}_{-}| \approx 1$, then one can also include the ``streaming losses'' (which reflect asymmetric scattering transferring energy into these rapidly-damped \Alf\ waves which then thermalize on short timescales, see  \citealt{Wien13,wiener:cr.supersonic.streaming.deriv,Rusz17,thomas.pfrommer.18:alfven.reg.cr.transport}) which for the assumptions above take the simple form: $\dot{e}_{\rm th,\,gas} \approx |v_{A}\,\nabla P_{\rm cr}|/3$.

\section{Numerical Implementation}
\label{sec:numerical}

Numerically, we can estimate $e_{{\rm cr},\,i}$, the value of $e_{\rm cr}$ at the center-of-mass location ${\bf x}_{i}$ of cell $i$, by summing Eq.~\ref{eqn:ecr.single.source} over all sources:
\begin{align}
\label{eqn:ecr.numerical} e_{{\rm cr},\,i} &\approx \sum_{j}\,\frac{\langle \dot{E}_{\rm cr} \rangle_{j}\,e^{-\tau_{{\rm cr},\,ij}}}{4\pi\,r_{ij}\,(\kappa_{0} + {v}_{\rm st,\,eff}\,r_{ij})} \\ 
& \rightarrow e^{-\Delta \tau_{{\rm cr},\,i}}\,\sum_{j}\,\left( \langle \dot{E}_{\rm cr} \rangle_{j}\,e^{-\Delta \tau_{{\rm cr},\,j}} \right)\,\mathcal{F}(r_{\rm ij}) 
\end{align}
i.e.\ 
\begin{align}
e_{{\rm cr},\,i} &\approx \mathcal{Q}^{\rm atten}_{i}\, \sum_{j}\, \dot{E}_{{\rm cr},\,j}^{\rm atten}\,\mathcal{F}(r_{ij})
\end{align}
with $r_{ij} \equiv |{\bf x}_{j} - {\bf x}_{i}|$, $\mathcal{Q}^{\rm atten}_{i} \equiv e^{-\Delta \tau_{{\rm cr},\,i}}$, $\dot{E}_{{\rm cr},\,j}^{\rm atten} \equiv Q^{\rm atten}_{j} \,\langle \dot{E}_{\rm cr} \rangle_{j}$. The source term $\langle \dot{E}_{\rm cr} \rangle_{j}$ is defined below, and the second equality stems from the following definitions and approximations: 
\begin{align}
\tau_{{\rm cr},\,ij} &\equiv \Delta\tau_{{\rm cr},\,i} + \Delta\tau_{{\rm cr},\,j}  + \left( \frac{r_{ij}}{r_{\rm max}} \right)^{2} \\ 
\Delta\tau_{{\rm cr},\,i}  &\equiv \frac{\psi_{\rm loss}^{i}}{2}\, \left[ \Delta x_{i}^{2} + \left( \frac{\rho_{{\rm gas},\,i}}{|\nabla \rho_{{\rm gas},\,i}|}  \right)^{2} \right]^{1/2} \\ 
\label{eqn:cr.egy.touse} \mathcal{F}(r_{ij}) &\equiv \frac{1}{4\pi\,r_{ij}\,(\kappa_{0} + v_{\rm st,\,eff}\,r_{ij})}\,e^{-r_{ij}^{2}/r_{\rm max}^{2}} 
\end{align}
Here we have approximated $\tau_{{\rm cr},\,ij}$ by the sum $\Delta \tau_{{\rm cr},\,i} + \Delta \tau_{{\rm cr},\,j}$, where $\Delta \tau_{{\rm cr},\,i}$ is (half, to account for the averaging) the integral $\tau_{\rm cr}$ extrapolated from the location of cell $i$ with a local Sobolev-length approximation, assuming a log-linear scaling $\rho(r)$ from $r=0$ to $r\rightarrow\infty$, based on the local gradient, giving the gradient scale-length $\rho/|\nabla \rho|$ evaluated at ${\bf x}={\bf x}_{i}$, plus the integral through the single cell $\Delta x_{i} \equiv (\Delta m_{i}/\rho_{i})^{1/3}$. This is akin to a LEBRON-type local radiation-hydrodynamics approximation \citep{hopkins:rad.pressure.sf.fb,hopkins:2019.grudic.photon.momentum.rad.pressure.coupling,hopkins:radiation.methods}. We average from both ``endpoints'' $i$ and $j$ to approximate the integral in-between.\footnote{This is convenient computationally as it reduces the computation to two local operations which can be done before and after the collective sum over sources, making the evaluation of Eq.~\ref{eqn:ecr.numerical} formally equivalent to the usual self-gravity evaluation. Integrating over the full ``path'' of a CR group to evaluate $\tau_{{\rm cr},\,ij}$ on the other hand would not only require evaluation of the integral of $\psi_{\rm loss}$ along the ray between source and target cell, but because the CR follows field lines and has a quasi-diffusive trajectory, we would have to integrate over the entire volume traversed by CRs (see \S~\ref{sec:assumptions}).}

We also impose the term $(r_{ij}/r_{\rm max})^{2}$, which accounts for finite-CR-transport time effects which are otherwise ignored in our calculation above which assumed that we were always in steady state ($\partial_{t} e_{\rm cr} \rightarrow 0$). In short, $r_{\rm max}$ reflects the maximum distance that CRs would travel in some time $t_{\rm max}$, so we do not inadvertently assign CR energy densities to gas arbitrarily far away from sources. For the same transport approximations we can calculate $r_{\rm max}$ numerically for some $t_{\rm max}$, or approximate it\footnote{Specifically, Eq.~\ref{eqn:rmax} is an approximate fitting function which is, by construction, asymptotically exact to the (already simplified) equations being integrated (e.g.\ Eq.~\ref{eqn:ecr.spherical} with constant $\kappa$, $v_{\rm st,\,eff}$) in the limits where either $\kappa$ or $v_{\rm st,\,eff}$ dominates (and the other can be neglected), and is within $\sim 10\%$ of the numerically-integrated solution for all values of interest of the dimensionless ratio $\kappa_{\rm iso}/v_{\rm st,\,eff}^{2}\,t_{\rm max}$. It represents the distance enclosing $1/2$ of the total energy of an initial $\delta$ function ``pulse'' of CRs injected at the origin, evolved according to Eq.~\ref{eqn:ecr.spherical}, after a time $t_{\rm max}$.} as 
\begin{align}
\label{eqn:rmax} r_{\rm max} &\equiv \frac{t_{\rm max}\,v_{\rm st,\,eff}}{2}\,\left[1 + \left( 1 + \frac{16\,\kappa_{\rm iso}}{v_{\rm st,\,eff}^{2}\,t_{\rm max}} \right)^{1/2} \right]
\end{align}
Note that this gives the correct behavior in both limits of $\kappa_{0}$ or $v_{\rm st,\,eff}$ dominating. The choice of $\tau_{\rm cr}$ scaling with $+(r_{ij}/r_{\rm max})^{2}$ ensures that in the diffusive limit, this reproduces exactly the Gaussian ``cutoff'' of the true constant-$\kappa_{0}$ finite-time diffusion solution. In cosmological simulations, it is reasonable to take $t_{\rm max} \approx t_{\rm Hubble}(z)$, the age of the Universe at redshift $z$, but one could also take it to be the time since the beginning of a simulation, or ideally the time since a source first formed or turned on (though this latter requires $t_{\rm max}$ for individual sources which can complicate implementation in tree methods, where one may need to define an ``effective'' average $t_{\rm max}$ for groups of sources which occupy the same tree node).

In Eq.~\ref{eqn:ecr.numerical}, approximating $\tau_{\rm cr}$ this way allows us to write $e_{{\rm cr},\,i}$ in a particularly useful form, after the $\rightarrow$ sign: we have a sum over all sources of a scalar, the ``locally attenuated'' $\dot{E}_{{\rm cr},\,j}^{\rm atten} \equiv \langle \dot{E}_{\rm cr} \rangle_{j}\,e^{-\Delta \tau_{{\rm cr},\,j}}$ (which can be evaluated and saved purely-locally for each source particle, as a single number that depends only on the local particle/cell properties), times a function $\mathcal{F}(r_{ij})$ which depends on their distance. And then after the sum is complete, we account for ``self-shielding,'' essentially by multiplying the saved sum at each site $i$ by $e^{-\Delta \tau_{{\rm cr},\,i}}$, in cell $i$. 

In tree codes, such as {\small GIZMO} which we test below, the implementation is then trivial and $e_{{\rm cr},\,i}$ can be computed with negligible cost alongside quantities like the gravitational forces. Specifically, all sources send one scalar, $\dot{E}_{{\rm cr},\,j}^{\rm atten}$, into the gravity force-tree, alongside the numbers needed for gravity, we perform the sum $\sum_{j}\,\dot{E}_{{\rm cr},\,j}^{\rm atten}\,\mathcal{F}(r_{ij})$ in the tree, using $r_{ij}$ between the target cell $i$ and all sources $j$ to evaluate $\mathcal{F}_{ij}$, then immediate after the tree-sum is complete multiply by the ``shielding'' term $e^{-\Delta \tau_{{\rm cr},\,i}}$, to obtain the updated $e_{{\rm cr},\,i}$ for that timestep. In the tree code, because the sum over $\dot{E}_{{\rm cr},\,j}^{\rm atten}$ is linear, we can make the same approximation as we do for gravity for distant nodes (replacing the individual $\dot{E}_{{\rm cr},\,j}^{\rm atten}$ with the sum in the tree node/branch/etc, and using the appropriate distance). It is immediately obvious how to generalize this to related multipole and other methods (although entirely Fourier-based methods require somewhat more overhead to correctly evaluate $\mathcal{F}(r_{ij})$). 

We can add the appropriate coupling terms to the gas, by inserting $e_{{\rm cr},\,i}$, and corresponding $P_{{\rm cr},\,i} = e_{{\rm cr},\,i}/3$, and their gradients, into all the existing Riemann problem and heating/cooling functions for the gas, exactly as described for our fully-explicit CR dynamics treatments in \citet{chan:2018.cosmicray.fire.gammaray}. 

All that remains is to specify the source terms $\langle \dot{E}_{\rm cr} \rangle_{j}$. Depending on the type of simulation, different approximations may be most useful. Most of the CR energy density in typical dwarf-through-MW-mass galaxies (at $\sim$\,GeV energies) is accelerated in SNe shocks, with $\epsilon_{\rm cr}^{\rm SNe} \sim 10\%$ of the ejecta kinetic energy ($E^{\rm SNe}_{\rm ej} \sim 10^{51}\,{\rm erg}$ per SNe) going into CRs, so $\langle \dot{E}_{\rm cr} \rangle_{j} \sim \epsilon_{\rm cr}^{\rm SNe}\,E^{\rm SNe}_{\rm ej}\,\langle \dot{N}_{\rm SNe}\rangle^{j}$ (where the average over $\dot{N}_{\rm SNe}$ should effectively ``smooth'' the SNe rate over some timescale e.g.\ of order the local dynamical time or a constant of $\gtrsim\,$Myr, to prevent spurious small-scale noise, since we assume steady-state solutions). In simulations which do not explicitly resolve young stellar populations (in time or space) the SNe rate and other forms of stellar feedback are commonly attached to star-forming gas-particles as sources, with $\langle\dot{N}_{\rm SNe}\rangle^{j} \sim (1/m_{\rm SNe,\,eff})\,\langle \dot{M}_{\ast} \rangle^{j}$, where $m_{\rm SNe,\,eff} \sim 100\,M_{\odot}$ depends on the assumptions about the stellar IMF and SNe progenitor mass range. In simulations which explicitly treat young stars and individual SNe, note that we still wish to have a continuum source for purposes of this CR sub-grid model (since it only applies in steady-state with some time-averaged injection rate), so we can take $\langle \dot{N}_{\rm SNe} \rangle^{j} \sim \langle R_{\rm SNe}(t_{\ast}^{j}) \rangle\,M_{\ast}^{j}$, where $R_{\rm SNe}$ is the SNe rate per unit stellar mass for a stellar population of age $t_{\ast}^{j}$ (used to compute all other stellar feedback effects in the code). If one wishes to include AGN as CR sources (reflecting relativistic jets), then one can take $\langle \dot{E}_{{\rm cr}}\rangle_{j} \sim \epsilon_{\rm cr}^{\rm BH}\,\langle \dot{M}_{\rm BH} \rangle^{j}\,c^{2}$, where $\dot{M}_{\rm BH}$ is the accretion rate onto the black hole in the AGN source (again, averaged over some smoothing time if the accretion model allows for arbitrarily short-timescale AGN variability, since we have assumed steady-state solutions) and $\epsilon_{\rm cr}^{\rm BH}$ parameterizes the fraction of the accretion energy which does into escaping relativistic particles. One could also add source terms for resolved shocks, if one wished to model structure formation shocks, for example, as CR sources.

Note that this adds no numerical timestep constraints, outside of those already present for MHD (if CRs modify the velocities or accelerations {\em of the gas}, the usual Courant conditions apply).

\begin{figure*}
	\includegraphics[width=0.96\textwidth]{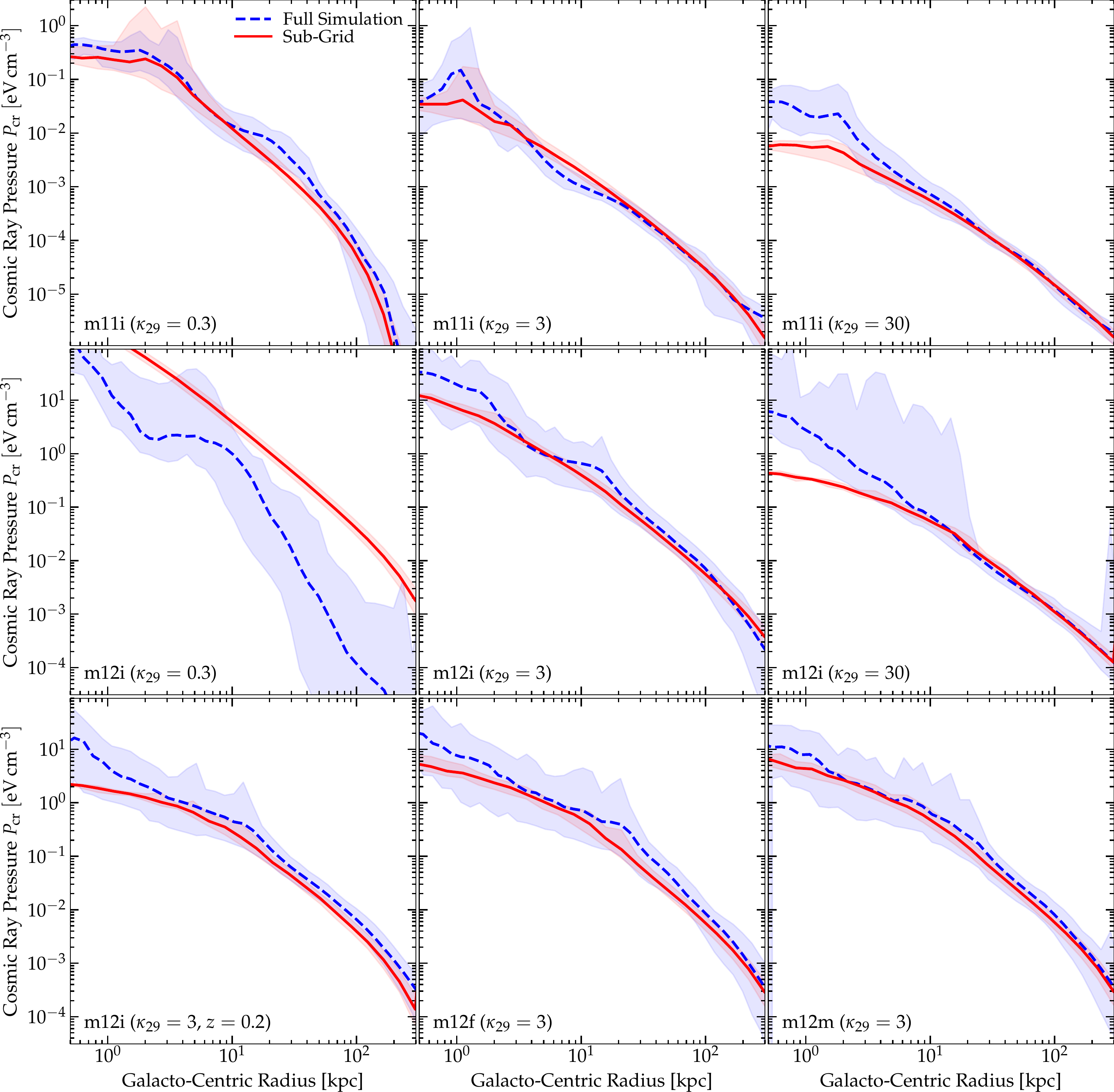}
	\vspace{-0.1cm}
	\caption{CR pressure $P_{\rm cr}$ (median [{\em solid}] and $\pm3\,\sigma$ range [{\em shaded}]) in spherical shells as a function of galacto-centric radius $r$, comparing simulations and sub-grid model predictions. ``Full simulation'' ({\em blue}) refers to values calculated for every gas cell in full cosmological numerical-hydrodynamic simulations which explicitly evolve CR transport including anisotropic diffusion, advection, variable streaming speeds, adiabatic losses/gains, and catastrophic losses (\S~\ref{sec:tests}). ``Sub-grid'' uses our simplified sub-grid model (\S~\ref{sec:numerical}) to estimate $P_{\rm cr} = e_{\rm cr}/3$. We compare different Milky Way-like galaxies at redshift $z=0$ ({\bf m12i}, {\bf m12f}, {\bf m12m})  and one dwarf (LMC-mass) galaxy ({\bf m11i}). The ``full simulations'' here assume a constant scattering rate $\bar{\nu}$, equivalent to a constant anisotropic/parallel diffusivity $\kappa_{\|} = c^{2}/3\bar{\nu}=\kappa_{29}\,10^{29}\,{\rm cm^{2}\,s^{-1}}$ plus streaming at the local \Alf\ speed $v_{A}$, and we compare simulations  with three different values of $\kappa_{29}$ spanning a reasonable range of observationally-allowed values. We also compare three different Milky Way-mass galaxies ({\bf m12f}, {\bf m12m}, {\bf m12i}), the same galaxy at different cosmological times ($z=0.2$, all others at $z=0$), and different galaxy masses ({\bf m11i}, an LMC-mass system). The sub-grid model has two free ``effective'' CR transport parameters ($\kappa_{0}$ and $v_{\rm st,\,eff}$): we calibrate these by fitting the profile to the $z=0$ snapshot of {\bf m12i} with $\kappa_{29}=3$ ({\em center} panel), then use these fitted values to extrapolate to all the other simulations shown here (assuming $v_{\rm st,\,eff}=$\,constant, and $\kappa_{0} \propto \kappa_{29}$). The sub-grid model can reasonably reproduce the CR pressure profile, especially in the CGM where it is most important. However for massive galaxies with high scattering rates (low diffusivities) compared to what is observed ({\bf m12i}, $\kappa_{29} \lesssim 0.3$), the sub-grid model substantially over-predicts the CR pressure because most of the CR energy in the full simulation is actually lost to catastrophic (pionic) processes, in contradiction to $\gamma$-ray observations.
	\label{fig:compare.profiles}}
\end{figure*}

\section{Assumptions}
\label{sec:assumptions}

Our sub-grid model makes many assumptions, which we have tried to enumerate in \S~\ref{sec:deriv}. Here we review which are ``reasonable'' or ``safe,'' and which are likely ``poor.''

To define ``good'' versus ``poor'' here: the key quantity of interest we wish to provide, for the community for whom this toy sub-grid model is intended and practically useful (e.g.\ large-volume galaxy formation simulations and semi-analytic models), is the cosmic ray pressure in the CGM (as for cosmological galaxy formation many studies have shown this can have some of the most dramatic effects, though it is far from the only way CRs can influence galaxy formation; see e.g.\ \citealt{Sale14cos} and the many references in \S~\ref{sec:intro}). But it is obvious (and supported by many detailed simulations and models; see e.g.\ \citealt{hopkins:2020.cr.transport.model.fx.galform,thomas:2022.self-confinement.non.eqm.dynamics} and references therein) that one can have reasonable models which behave (by construction) identically in the ISM and yet give order-of-magnitude differences in the CR pressure in the CGM, based on different assumptions about how to extrapolate the (deeply uncertain) CR transport parameters (e.g.\ $\kappa$, $v_{\rm st}$, etc.) to different plasma conditions. In principle these might be testable in the future (see discussion in \citealt{butsky:2022.cr.kappa.lower.limits.cgm}), but at present there are few observational constraints at the scale of interest (assuming the model is constrained at the order-of-magnitude level already in the ISM by Solar system and $\gamma$-ray data). This means that -- for our limited purposes here -- a ``good'' or ``safe'' or ``reasonable'' assumption in our toy model is one which, if changed, would not change the average CR pressure on some scale by multiple orders-of-magnitude, in a way that {\em could not} be absorbed into the already explicitly-parameterized uncertainties (e.g.\ parameters like $\kappa$ or $v_{\rm st}$) of the model. This, of course, is a much looser criterion than ``could in principle be observationally measured,'' let alone ``can we justify the assumption formally or rigorously.'' But it is the case of practical interest for the (intentionally extremely-simplified) models here.

\subsection{Well-Justified Assumptions} 

First, some assumptions we make are likely can actually be formally justified, as e.g.\ certain dropped terms are small compared to other terms we retain. As shown in \citet{hopkins:m1.cr.closure}, for scales $\gg r_{\rm g}$ (the CR gyro radius $\sim 0.1\,{\rm au}$) of interest, with non-relativistic MHD fluid motions, the gyrotropic expansion and expansion to leading-order in $\mathcal{O}(u/c)$ are well-motivated (i.e.\ CR gyro radii are small compared to the scales of application of the model, and the background fluid motions are non-relativistic; see e.g.\ \citealt{jokipii:1966.cr.propagation.random.bfield,skilling:1971.cr.diffusion}). Likewise on scales $\gtrsim$\,kpc much larger than the CR scattering mean-free-path, since CRs are not truly ``collisionless'' like photons, the moments approach to CR dynamics and assumption of a near-isotropic CR distribution function and flux-steady state are also formally justified \citep{1975RvGSP..13..547V}. And as shown in \citet{hopkins:cr.multibin.mw.comparison}, for realistic diffusivities and CR spectra, neglecting terms such as the re-acceleration (though we discuss this further below) and certain other losses (e.g.\ ionization for low-energy protons, or losses for sub-dominant leptonic CRs) are also likely to be a very small source of error (they will at least always be smaller than other terms we {\em also} neglect, such as the adiabatic term, so it is more important to discuss those terms in our breakdown below). 

\subsection{Weakly-Justified, But Plausible Assumptions on Galactic Scales}

Next, we have a group of assumptions which are not rigorously motivated, but work surprisingly ``well'' in the loose practical sense we define above, if we focus on scales $\sim 1-100\,$kpc around galaxies. This includes the following.

We integrate over CR momentum to use spectrally-integrated CR equations assuming the ultra-relativistic limit, which assumes the CR spectrum is self-similar. This usually works at an ``acceptable'' level because most of the CR energy is around $\sim1-10\,$GeV even if the spectral shape varies (except perhaps near the galactic center; see \citealt{Sale16,Chen16,Buts18,chan:2018.cosmicray.fire.gammaray}), and we only care about CR effects on galaxies (dominated by the total pressure, without strong dependence on CR spectral shape) and not CR observables (where the spectral shape is very important). The spectral shape variations would have to be extreme to change the total CR pressure at the order-of-magnitude level, and even if these did appear, since in any realistic model the CR transport parameters depend on rigidity, it is not obvious it could not be subsumed into an appropriate mean radial dependence of $\kappa$ or $v_{\rm st}$ on $r$.

We assume magnetic fields are ``tangled'' so we can approximate the diffusivity as isotropic, for analytic simplicity. This will break down badly on small scales, of course, but is plausible on large/CGM scales (see \citealt{ji:fire.cr.cgm,ji:20.virial.shocks.suppressed.cr.dominated.halos}). More importantly however, multiple previous studies have shown that this generally only introduces $\mathcal{O}(1)$ geometric corrections to the effective transport speed and CR pressure even where magnetic fields are highly-aligned (\citealt{Chen16,buck:2020.cosmic.ray.low.coeff.high.Egamma,hopkins:cr.mhd.fire2,wehahn:2021.gamma.rays,ji:fire.cr.cgm,butsky:2022.cr.kappa.lower.limits.cgm}; akin to other anisotropic diffusion processes; see \citealt{sharma.2010:thermal.anisotropic.conduction.instability.clusters,parrish.2012:anisotropic.viscosity.bouyancy.instability.effects,arth.2014:anisotropic.conduction.sph.gadget,hopkins:gizmo.diffusion}), sub-dominant compared to other effects and easily absorbed into the ``effective transport parameters.'' Most noteworthy, the simulations we compare/calibrate to here all include strictly anisotropic transport along magnetic field lines, so this calibration is already automatically included in our analysis.

We also neglect the ``streaming losses'' at large CGM radii. This could in principle be a less-accurate assumption if the effective streaming  speed at these radii drops from highly super-\Alf{ic} (required near/within the galaxy) to strictly locked to the \Alf\ speed and the \Alf\ speed in the halo were very low, but recall we have assumed constant transport coefficients so this should not occur (though it could if one adopted very different transport models). And the observationally-required values of the diffusivity/streaming speed within galaxies are so large that if they are constant, this introduces negligible error; moreover in the CGM it is extremely difficult in simulation models for this loss term to strongly suppress the pressure when various effects including out-of-equilibrium transport are considered \citep{thomas:2022.self-confinement.non.eqm.dynamics}. But even the ``worst-case'' version of this is unlikely to introduce order-of-magnitude reductions in the CR pressure.

We use a LEBRON-type two-endpoint approximation for the ``attenuation'' $\tau_{{\rm cr},\,ij} \approx \Delta \tau_{{\rm cr},\,i} + \Delta \tau_{{\rm cr},\,j}$ of CRs. This could in principle miss ``shadowing'' effects by dense clumps in-between the CR sources and gas cells. In RHD the accuracy of this approximation can be improved (at greater computational expense) by replacing the two-endpoint approximation with a HealPix/TreeCol-type method or a full ray-tracing integration, but that assumes photons travel from sources on ``straight-line'' (ray) trajectories until absorbed. But since the CRs are actually scattering/diffusive and moving along tangled fields, this would not actually improve the accuracy significantly, and they are never in the truly ``optically thin'' (aka ``negligible scattering'') equivalent limit. Our approximation captures the fact that for losses to strongly deplete the CRs, or shielding strongly suppress their effects, it must happen {\em quasi-isotropically} owing to the quasi-isotropic scattered CR trajectories near to the source or ``target.'' Most importantly again, the simulations here to which we calibrate do not make this approximation, and yet we see only modest inhomogeneity actually appear at large radii, most of it owing not to ``shadowing'' but to the adiabatic terms discussed below, so it can be incorporated into our calibration of the models fairly easily.

Spherical symmetry is obviously a poor approximation on small scales, but again, because CRs are ``diffusive'' and scatter, especially if fields are tangled, then CRs sample and implicitly smooth over an entire volume as they escape (and again, ``shadowing'' or other ``optically-thin-like'' RHD limits, do not apply to CRs). This means that even in an inhomogeneous medium, the effective steady-state solution at a given point reflects some weighted average over the whole volume, suppressing the effects of anisotropy and producing quasi-spherical CR profiles (with only tens of percent deviations) in full explicit CR-dynamics simulations \citep{Buts18,hopkins:2020.cr.outflows.to.mpc.scales,ji:20.virial.shocks.suppressed.cr.dominated.halos}. Thus especially on CGM scales of greatest interest in many CR studies, this is a surprisingly ``acceptable'' approximation as far as our (loose) criteria are concerned.

\subsection{Most Uncertain Assumptions}
\label{sec:assumptions.uncertain}

The assumptions which likely drive the largest uncertainties (but are necessary for the simplicity of our model) on galactic scales include the following.

We assume steady-state for the CR energy equation and continuous injection, which clearly breaks down on small scales, where finite-source time and space structure (e.g.\ injection in super-bubbles and non-equilibrium effects given that the CR escape time is comparable to galaxy and GMC dynamical times) will be important. On very large scales, finite-travel time effects mean energy steady-state cannot hold, as the diffusion time $\sim \ell^{2}/\kappa \sim (0.3\,{\rm Gyr})\,(\ell/10\,{\rm kpc})^{2}\,(\kappa/10^{29}\,{\rm cm^{2}\,s^{-1}})$ becomes long. This is why we add the ``$r_{\rm max}$'' term in our numerical method, to at least partially account for this. But the simple fact is that on large/CGM scales, it is quite possible that the CR properties are far from steady-state. This is especially a concern if one wishes to model CR injection from AGN, where a bright quasar could introduce an enormous instantaneous CR injection rate, but this would last for a relatively short time, so the dynamics would resemble more of a ``pulse'' of CRs propagating outwards (which could easily require multiple Gyr to reach the virial radius for the typical transport parameters invoked here), rather than a steady-state solution.

We neglect adiabatic/convective terms. These generally have an order-unity effect on CR energies, and their sign can vary (enhancing CR densities in dense, collapsing gas, and decreasing it in outflows; \citealt{2017ApJ...847L..13P,chan:2018.cosmicray.fire.gammaray}). This means also that we will not capture the full local variation in e.g.\ the CR ionization rate, an issue exacerbated by the fact that we do not treat a full CR spectrum, so neglect the fact that low-energy CRs are more tightly-coupled to the gas \citep{hopkins:cr.multibin.mw.comparison}. And it means we cannot capture certain effects of CRs modifying the properties of local thermal instabilities, in the ``partially-coupled'' regime \citep{butsky:2020.cr.fx.thermal.instab.cgm}. From the analysis in \citet{chan:2018.cosmicray.fire.gammaray} and \citet{buck:2020.cosmic.ray.low.coeff.high.Egamma}, this appears to contribute significantly to the local (small-scale) fluctuations in the CR pressure seen in the full simulations at large radii, in a way that our sub-grid models simply do not capture (they necessarily predict much less scatter because of the neglect of these local dynamics). So while unlikely to change things at the order-of-magnitude level, this could bias the results (both in terms of the effects of CRs for a given diffusivity, or as we show below the inferred ``best-fit'' diffusivity to full dynamical simulations) by as much as a factor of a few, via non-linear interaction channels in particular such as the effect of ``partial'' adiabatic coupling on the non-linear evolution of the thermal instability in the CGM \citep{butsky:2020.cr.fx.thermal.instab.cgm}, and its non-linear effect in turn on accretion and clumpiness of the CGM \citep{ji:fire.cr.cgm}. 

Finally, from a CR physics point of view, likely by far the least-accurate assumption we make is to assume that the CR transport coefficients are constant in space and time. In physically-motivated models of CR transport, whether the scattering comes from self-confinement (SC) or extrinsic turbulence (ET), the prediction is that the scattering-rate coefficients should be strong functions of local plasma properties such as the magnetic field strength, \Alf\ Mach number and dissipation rate of the local turbulence, gas density, temperature, ionized/neutral fractions, dust-to-gas ratio, and others \citep[see e.g.][]{chandran00,yan.lazarian.02,yan.lazarian.04:cr.scattering.fast.modes,yan.lazarian.2008:cr.propagation.with.streaming,farmer.goldreich.04,zweibel:cr.feedback.review,thomas.pfrommer.18:alfven.reg.cr.transport,hopkins:cr.transport.constraints.from.galaxies,squire:2021.dust.cr.confinement.damping}. This leads to scattering rates (hence transport coefficients) which can vary by orders-of-magnitude even within a $\sim$\,kpc ``patch'' of the ISM \citep[for observational evidence, see][]{hawk:2017.pwne.apparent.low.cr.diffusion.in.nebula,hess:2019.excess.around.cosmic.ray.pulsar.wind.nebula.low.diffusion.maybe}, and the mean spherically-averaged scattering-weight-weighted transport parameters can vary by orders-of-magnitude systematically as a function of galacto-centric radius \citep{hopkins:cr.transport.constraints.from.galaxies,hopkins:2020.cr.transport.model.fx.galform}. But there are two problems if we wish to devise a sub-grid model to represent these physics. First, it is more difficult to design a simple and low-cost sub-grid model if the CRs scatter through a field with rapidly-varying local diffusivity/streaming speed, since even some appropriate ``average'' must be weighted over the entire  scattering domain (i.e.\ we must know about all points along all possible scattering trajectories, not just the ``line of sight''). But second and more challenging, it is not at all clear which (if any) of the models for CR scattering rates surveyed in \citet{hopkins:cr.transport.constraints.from.galaxies} are correct, and these models disagree {\em qualitatively} in the scalings with these local properties (even the qualitative sense of whether scattering rates should increase or decrease with larger magnetic field strengths or turbulent dissipations rates remains controversial). Moreover, \citet{hopkins:2021.sc.et.models.incompatible.obs} show that the textbook \citep{jokipii:1966.cr.propagation.random.bfield,kulsrud.1969:streaming.instability} and state-of-the-art \citep{yan.lazarian.04:cr.scattering.fast.modes,thomas.pfrommer.18:alfven.reg.cr.transport} theories for scattering from ET or SC all fail qualitatively in predicting Solar system CR properties at $\sim$\,GeV energies, generically predicting orders-of-magnitude too-small or too-large $\kappa$ (or $v_{\rm st}$), and worse yet predicting that the CR scattering rates and residence times should actually be constant or {\em increase} with CR rigidity from $\sim\,$GV-TV (the opposite of the observed behavior).\footnote{These behaviors have also been noted for SC going back to at least \citet{cesarsky.kulsrud:1981.cr.confinement.damping.hot.gas} and for ET in \citet{chandran00,yan.lazarian.2008:cr.propagation.with.streaming}. As reviewed in \citet{hopkins:2021.sc.et.models.incompatible.obs} and \citet{kempski:2021.reconciling.sc.et.models.obs}, they relate fundamentally in SC to the fact that the CR distribution function appears in the effective ``diffusion coefficient,'' rapidly forcing the system towards highly-constrained equilibrium solutions, while for ET, they relate to the basic mathematical structure of the MHD equations coupled to the anisotropy of basic plasma processes such as Landau damping which become important for CRs at energies below $\sim$\,TeV-PeV. Possible resolutions to these issues from a plasma-physics CR scattering theory perspective are reviewed in both \citet{hopkins:2021.sc.et.models.incompatible.obs} and \citet{kempski:2021.reconciling.sc.et.models.obs}.} As such, ``constant diffusivity'' or ``constant-$v_{\rm st,\,eff}$'' approximations, while almost certainly incorrect, remain popular by necessity. Our intention with the simplified models presented here it to be agnostic to these uncertainties -- to, as much as possible, allow users to insert their own assumptions and use these to attempt to constrain reasonable ``effective transport parameters'' constrained as very large-scale effective averages (on e.g.\ scales of the virial radii of galaxies, for example).

\begin{figure}
	\includegraphics[width=0.95\columnwidth]{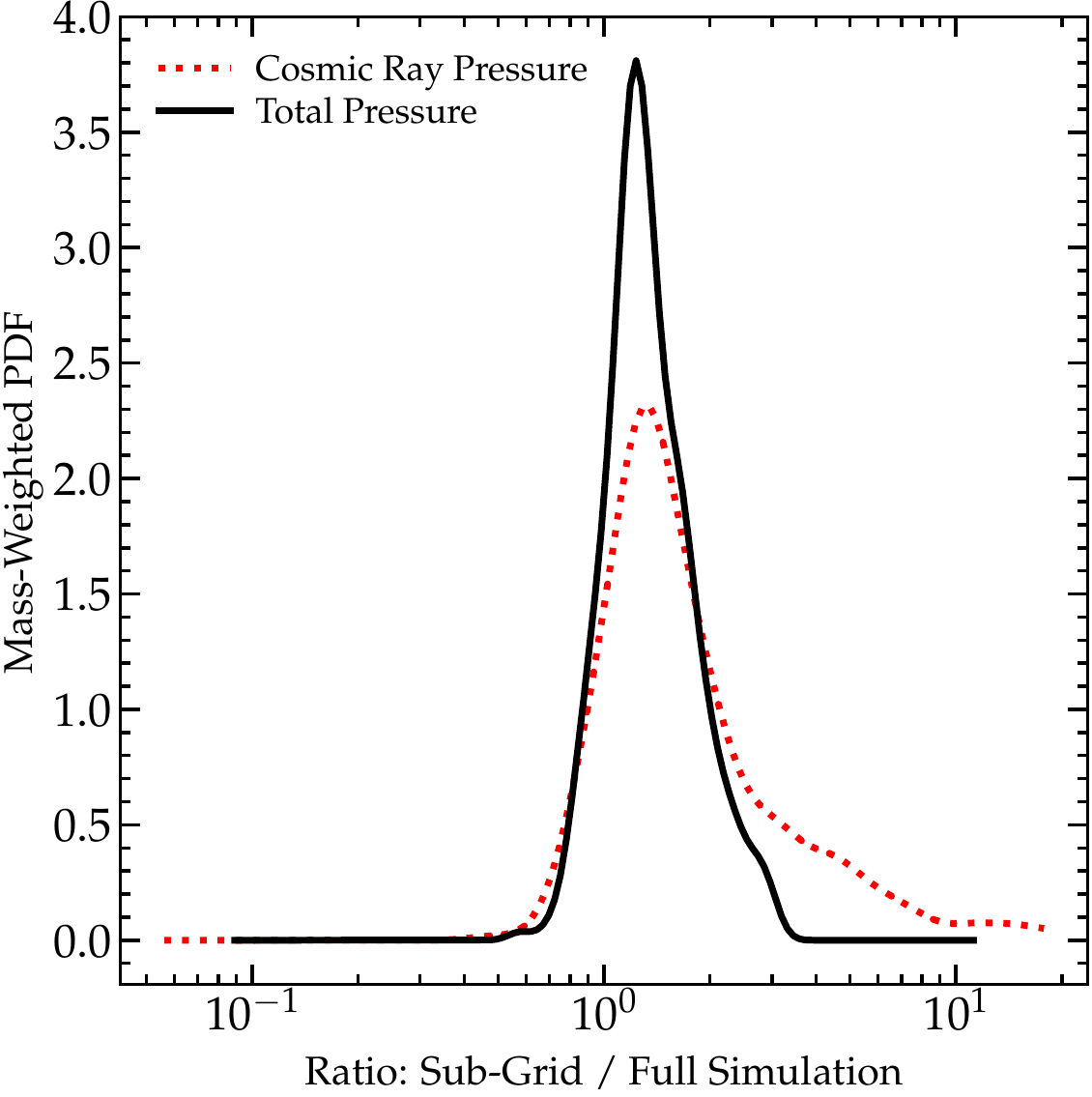}
	\vspace{-0.1cm}
	\caption{Distribution over all gas cells (weighted by gas mass) of the sub-grid estimated pressure to the actual simulation pressure, for the {\bf m12i}, $\kappa_{29}=3$ simulation in Fig.~\ref{fig:compare.profiles}. We compare both the ratio of CR pressure specifically, $P_{\rm cr}^{\rm subgrid} / P_{\rm cr}^{\rm \,simulation}$ ({\em red dotted}) and the ratio of total pressures $P_{\rm tot}^{\rm subgrid} / P_{\rm tot}^{\rm \,simulation}$ ({\em black solid}). In a gas-mass-averaged sense, the CR pressure is generally estimated with factor $\sim 2$ scatter, and the tail of cases where the sub-grid model for $P_{\rm cr}$ is wrong by a large factor generally corresponds to cases where the CRs contribute little to the total pressure.
	\label{fig:pdf}}
\end{figure}

\section{Validation In Numerical Simulations}
\label{sec:tests}

\subsection{Analysis Sample}
\label{sec:tests:sims}

We now validate this sub-grid model by comparing its predictions for CR pressure to the results of fully-explicit CR-MHD galaxy-formation simulations. The simulations are presented in \citet{hopkins:cr.mhd.fire2}, run with {\small GIZMO}\footnote{A public version of {\small GIZMO} is available at \gizmourl} \citep{hopkins:gizmo,hopkins:mhd.gizmo,hopkins:cg.mhd.gizmo} as part of the Feedback In Realistic Environments (FIRE)\footnote{\FIREurl} project \citep{hopkins:2013.fire,hopkins:fire2.methods}. These are fully-cosmological, hight-resolution MHD simulations which follow the formation of galaxies from initial conditions at $z\gtrsim 100$ to $z=0$ and include explicit treatment of multi-phase gas thermo-chemistry and radiative cooling, star formation, and stellar feedback from the stars that form in a variety of forms including mechanical (from stellar mass-loss, core-collapse and Ia supernovae) and radiative (including radiation pressure and various photo-heating and ionization terms), as well as CRs (with each individually time-resolved SNe injecting $\epsilon_{\rm cr}^{\rm SNe}\sim10\%$ of the initial ejecta kinetic energy into CRs). The CR transport is evolved fully-explicitly, with all relevant gas coupling terms in the original simulations, following the spectrally-integrated CR energy and flux equations allowing for fully-anisotropic diffusion and streaming, with $\bar{\nu}=$\,constant\footnote{A constant scattering rate $\bar{\nu}$ roughly translates to a constant parallel/anisotropic diffusivity $\kappa_{\|} \approx c^{2}/3\,\bar{\nu}$, but we stress that the diffusivity emerges as a function of $\bar{\nu}$ and $\bar{v}_{A}$ from the explicitly-evolved CR equations when the CR flux equation approaches local quasi-steady-state.} and $\bar{v}_{A}=v_{A}$. These are useful because the uncertainty in true locally-variable CR scattering rates, which is a physical uncertainty beyond the scope of our sub-grid model, does not enter here. 

However, we will also consider a subset of simulations from study in \citet{hopkins:cr.transport.constraints.from.galaxies,hopkins:2020.cr.transport.model.fx.galform}. These simulations adopt identical initial conditions and numerical methods, except that the scattering rates (hence parallel diffusion coefficient and streaming speed) are allowed to be complicated functions of the local plasma parameters (e.g.\ ionization state, density, magnetic field strength, turbulence, etc.), motivated by theoretical self-confinement and extrinsic turbulence models for CR scattering. We will restrict our comparisons to the subset of models which \citet{hopkins:cr.transport.constraints.from.galaxies} showed could be possibly consistent with existing CR observations, studied in more detail in \citet{hopkins:2020.cr.transport.model.fx.galform}.

In either case, we post-process snapshots from these simulations with our sub-grid model calculating $e_{\rm cr}$ (or $P_{\rm cr}$) exactly as we would ``on the fly'' per Eq.~\ref{eqn:ecr.numerical}, and compare it to the true values given in-code from the explicit CR-MHD dynamics simulations.

\begin{figure*}
	\includegraphics[width=0.99\textwidth]{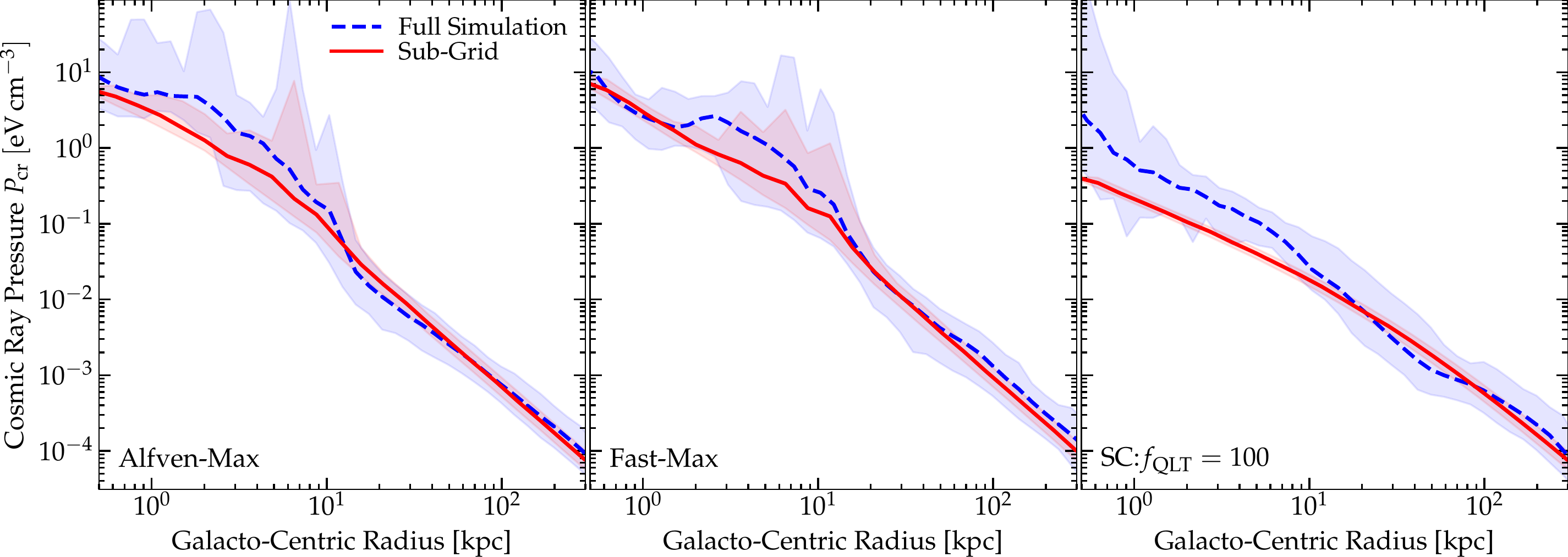}
	\vspace{-0.1cm}
	\caption{As Fig.~\ref{fig:compare.profiles}, but comparing simulations of galaxy {\bf m12i} from \citet{hopkins:cr.transport.constraints.from.galaxies} which adopt locally dynamically-variable CR transport (diffusion/scattering/streaming) coefficients which are complicated functions of local plasma properties in the ISM, motivated by more physical self-confinement or extrinsic turbulence theories for CR scattering. {\bf Alfven-Max}, {\bf Fast-Max}, and {\bf SC:$f_{\rm QLT}=100$} represent models (normalized to fit Solar system CR observations) where scattering rates specifically for $\sim 1\,$GeV protons scale with ISM plasma properties dimensionally as expected from \Alf{ic} turbulence, magnetosonic turbulence, or self-confinement via gyro-resonant instabilities. Here we re-calibrate the sub-grid model parameters $\kappa_{0}$ and $v_{\rm st,\,eff}$ for each simulation, as they feature very different physics and ``mean'' effective coefficients; however we focus on a subset of models which are at least conceivably consistent with Milky Way CR observations as discussed in \citet{hopkins:cr.transport.constraints.from.galaxies}. With this constraint, it appears that the allowed range of ``effective'' $\kappa_{0}$ and $v_{\rm st,\,eff}$ is relatively modest, approximately following Eq.~\ref{eqn:kappa.constraint}, and the average profiles can still be reasonably represented by our toy model (for some appropriate ``effective'' $\kappa_{0}$, $v_{\rm st,\,eff}$) despite the true local scattering rates varying by large factors.
	\label{fig:compare.profiles.variable}}
\end{figure*}

\subsection{Results: Post-Processing Comparisons}
\label{sec:tests:results}

\subsubsection{Calibration}
\label{sec:tests:results:calibration}

Fig.~\ref{fig:compare.profiles} compares the radial profiles of the CR pressure calculated from our simple sub-grid model and the explicit-CR-dynamics simulations. Our initial reference case (halo {\bf m12i} with constant $\bar{\nu} = 10^{-9}\,{\rm s^{-1}}$ (equivalent to a local-steady-state anisotropic $\kappa_{\|} = c^{2}/3\,\bar{\nu} = 3\times10^{29}\,{\rm cm^{2}\,s^{-1}}$) is one shown in \citet{hopkins:cr.mhd.fire2,hopkins:cr.transport.constraints.from.galaxies} to be both a reasonable Milky Way analogue in its galaxy properties and to reproduce (for the chosen $\bar{\nu}$ or $\kappa_{\|}$) various observational constraints on CRs including $\gamma$-ray emission, grammage, and CR lifetime constraints from secondary-to-primary and radioactive isotope ratios (see references in \S~\ref{sec:tests:results:obs.input}).\footnote{As we discuss below, when comparing this ``fiducial'' diffusion coefficient $\kappa$ in the simulations to typical quoted values in the CR literature of e.g.\ $D_{xx}$, it is important to account for three things.(1) Since we integrate over the entire CR spectrum, $\kappa$ here should be an energy-weighted average $\langle D_{xx} \rangle$ over the CR spectrum, which boosts $\langle D_{xx} \rangle$ by a factor of several over $D_{xx}[{\rm R=GV}]$. (2) The $\kappa$ we quote is the parallel/anisotropic coefficient, while $D_{xx}$ refers to an isotropically-averaged coefficient, which is $\sim 3\times$ lower than $\kappa$. And (3) the ``effective'' $\kappa$ we quote also (by definition) includes any advective/convective and/or streaming velocities, which are often quoted or fit separately in classic models for some $D_{xx}$. Considering effects (1)-(3) above, we can compare the best-fit models from recent studies like \citet{evoli:2019.cr.fitting.galprop.update.ams02} or \citet{korsmeier:2022.cr.fitting.update.ams02} which fit $D_{xx}[{\rm R\sim GV}]\sim 10^{28}\,{\rm cm^{2}\,s^{-1}}$ (combining with the fits to the full CR spectrum from \citealt{bisschoff:2019.lism.cr.spectra} to perform the spectral integration) and obtain the corresponding ``effective'' $\kappa \approx 3-5\times 10^{29}\,{\rm cm^{2}\,s^{-1}}$ for the best-fit models therein. This is remarkably close to our ``fiducial'' simulation choice (which is of course part of the reason we refer to it as ``fiducial'').}
However we stress that our modeling infrastructure is agnostic to the actual transport parameters and appears to work similarly well for a range of observationally-allowed values. Explicitly, the sub-grid model has two ``free'' parameters: $\kappa_{0}$ and $v_{\rm st,\,eff}$, which represent physical terms but must be calibrated at some level to the full simulations. Even for a simulation like this reference case where the in-code scattering rate or parallel diffusion coefficient is taken to be a constant, recall that our sub-grid $\kappa_{0}$ represents an isotropically angle-averaged ``diffusivity,'' neglecting terms such as advection and adiabatic CR gains/losses and out-of-flux-equilibrium drift terms, treating CRs as if we can add sources independently, and further assuming spherical symmetry around each source (and the effects and accuracy of  all these approximations will vary in both space and time). Likewise $v_{\rm st,\,eff}$ implicitly represents all of these approximations plus, in the simulation, the actual streaming speed (the \Alf\ speed), which is locally-variable. So it is by no means obvious, a priori, that our subgrid model will be able to fit the simulation results for {\em any} constant $\kappa_{0}$ and $v_{\rm st,\,eff}$. 

Nonetheless, Fig.~\ref{fig:compare.profiles} shows that we can obtain a quite reasonable match to the simulation profile if we adopt $\kappa_{0} \approx 5\times10^{28}\,{\rm cm^{2}\,s^{-1}}$ and $v_{\rm st,\,eff} \approx 20\,{\rm km\,s^{-1}}$. These are plausible values: if there were no other approximations in our model but we simply had isotropically tangled fields, then the isotropically-averaged diffusivity in the simulation would be $\kappa_{\rm iso} \approx \kappa_{\|}/3 \approx 10^{29}\,{\rm cm^{2}\,s^{-1}}$ -- just a factor of $\sim 2$ larger than $\kappa_{0}$ we infer by essentially fitting the simulation profile. The difference owes to the (many) other assumptions reviewed in \S~\ref{sec:assumptions}. Notably, since the medium on small scales (within/near the galaxy) is not really spherically symmetric with a single point source at the center, the profile has a higher ``flat shelf'' in the center than our toy sub-grid model would predict,  which leads us to fit a slightly lower ``effective'' diffusivity. This can be illustrated by simply considering two point sources (where our toy model would predict a local maximum between the two, but in reality this would be diffused out). This also leads to some ``tradeoff'' where in our toy model we tend to fit slightly lower $\kappa_{0}$ and higher $v_{\rm st}$ than were present in our original simulation: here, while $\sim20\,{\rm km\,s^{-1}}$ is a plausible mean \Alf\ speed $v_{A}$, recall this is isotropically averaged so should be a factor $\sim 3$ lower, if we were recovering exactly the input simulation parameters. This demonstrates the importance of using ``full'' simulations like those here to calibrate these sub-grid models, before applying them more broadly.

With these choices, we can compare not just the radial trend of CR energy density, but also the values cell-to-cell (i.e.\ the scatter or difference therein in the curves in Fig.~\ref{fig:compare.profiles}. Specifically, Fig.~\ref{fig:pdf} shows the distribution of values of $P_{\rm cr}$ inferred from our sub-grid model compared to those in the live simulation. This works remarkably well, reproducing the simulations to within a factor $\sim 2$ scatter.  Note that when $P_{\rm cr}$ is very small, even a tiny inaccuracy in the sub-grid model could lead to an apparently large fractional discrepancy ($|\ln{(P_{\rm cr}^{\rm sim}/P_{\rm cr}^{\rm subgrid})}| \gg 1$), but this would be irrelevant for the gas dynamics. We therefore also consider the effect on the total pressure, adding the CR pressure to the thermal and magnetic pressures in code. This allows us to see that the ``tail'' of cases where the sub-grid model fails badly and in estimating $P_{\rm cr}$ are mostly cases where the CR pressure is irrelevant -- so these will not produce significant effects on the simulation dynamics. From comparison with Fig.~\ref{fig:compare.profiles}, we can immediately see that the scatter in Fig.~\ref{fig:pdf} is essentially equivalent to the scatter in the ``full'' simulation in $P_{\rm cr}$ at a given galacto-centric radius. By construction, the toy model here represents only the spatially spherically-averaged, and cosmic-ray-transport-time time-averaged value of $P_{\rm cr}$ at a given radius, so (as expected) exhibits very little scatter in $P_{\rm cr}$ at a given radius. Physically, the scatter in the full simulations arises from violations of homogeneity and spherical symmetry (e.g.\ winds, local sources, satellite galaxies, regions/phases of the ISM/CGM with rapid loss rates, regions where the local \Alf\ speeds are much higher, and the ``adiabatic terms'' -- all effects our toy model averages out. The most important effect of this scatter which our model does not capture is likely an indirect effect: namely, that in the CGM, the scatter is largely related to variations in the adiabatic term (see references in \S~\ref{sec:assumptions.uncertain}, who show this explicitly), because the CR diffusion/transport times are not necessarily extremely short compared to other bulk timescales on such large scales ($\sim 10-300\,$kpc). As shown in \citet{butsky:2020.cr.fx.thermal.instab.cgm}, this leads to non-trivial differences in the behavior of the thermal instability of the CGM (the toy model here is closest to the ``decoupling limit'' considered in \citealt{butsky:2020.cr.fx.thermal.instab.cgm}, while the simulations lie closest to one of the ``intermediate'' cases per \citealt{ji:fire.cr.cgm}). It is worth thinking, in future work, of sub-grid models which might be able to (partially) capture this effect via introduction of a more complex ``hybrid'' non-linear equation-of-state and cooling function in the gas that incorporates the transport parameters as well as the total CR pressure and gas thermo-chemistry.

\begin{figure}
	\includegraphics[width=0.95\columnwidth]{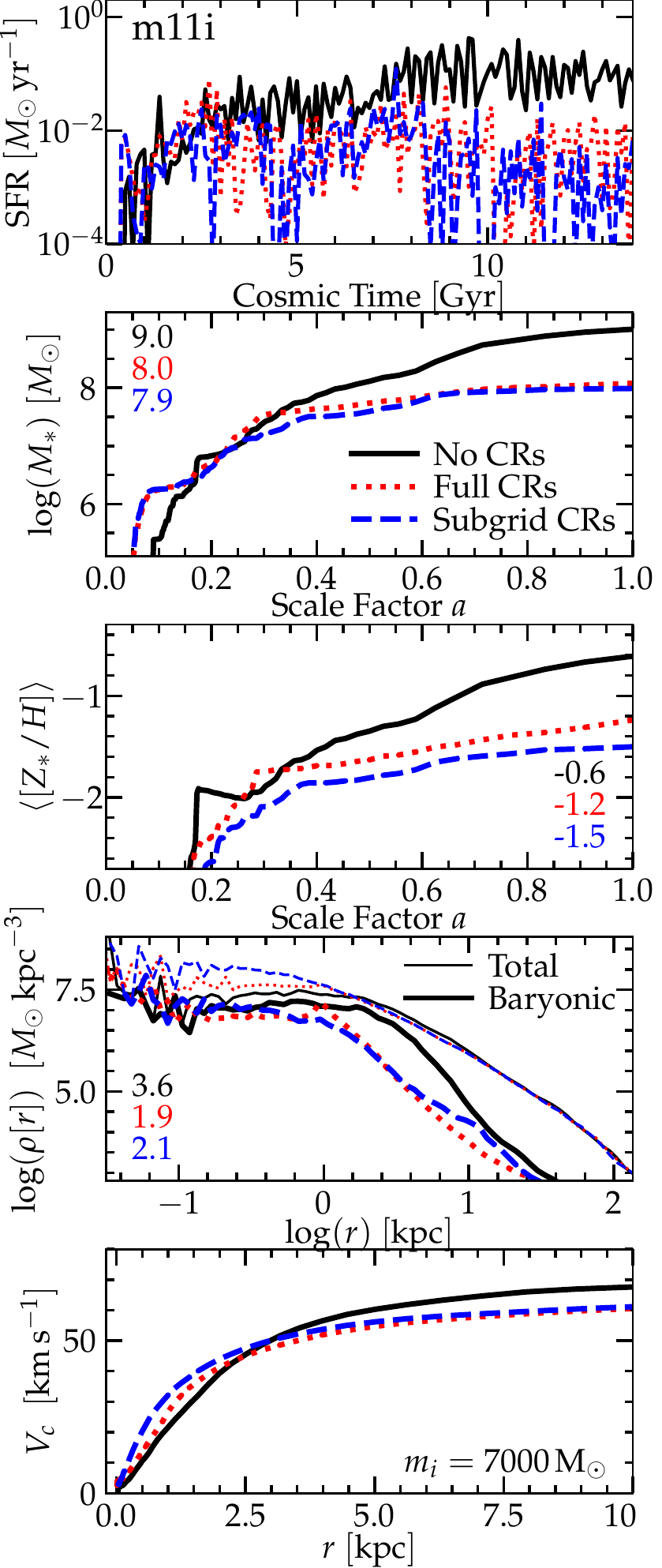}
	\vspace{-0.1cm}
	\caption{Example application of the sub-grid model in a cosmological simulation of {\bf m11i} (an LMC-mass system from Fig.~\ref{fig:compare.profiles}) from $z\sim100$ to $z=0$. We compare three physics variations: (1) with no CRs or MHD; (2) with the full explicit single-bin CR dynamics, gas coupling, and kinetic MHD from \citet{hopkins:cr.mhd.fire2}, with parallel (anisotropic) $\kappa_{\|}=3\times10^{29}\,{\rm cm^{2}\,s^{-1}}$ and streaming at the \Alf\ speed; (3) a run with no explicit CR dynamics or MHD, but adding our proposed sub-grid model, with constant $\kappa_{0}=0.5\times10^{29}\,{\rm cm^{2}\,s^{-1}}$, $v_{\rm st,\,eff}=20\,{\rm km\,s^{-1}}$, calibrated as  \S~\ref{sec:tests:results:calibration}. We compare:
	{\em Top:} Archeological SFR vs.\ cosmic time (averaged in $\sim100\,$Myr intervals).
	{\em Second:} Stellar mass vs.\ scale factor $a$. Numbers give the value $\log{(M_{\ast}/M_{\odot})}$ at $z=0$.
	{\em Third:} Mean stellar metallicity vs.\ $a$. Numbers give $\langle [ Z_{\ast}/H ] \rangle$ at $z=0$.
	{\em Fourth:} Spherically-averaged radial density profile of all mass (including dark-matter; {\em thin}) and baryonic ({\em thick}) at $z=0$, versus galacto-centric radius $r$. Numbers give the stellar effective radius $R_{\rm eff,\,\ast}$ in kpc at $z=0$.
	{\em Bottom:} Circular velocity ($V_{\rm c}^{2} \equiv G\,M_{\rm tot}(<r)/r$) profile. 
	 The subgrid model appears to reasonably capture the zeroth-order (but not all) effects of CRs on galaxy formation. 
	\label{fig:cosmo.test}}
\end{figure}

\subsubsection{Validation}
\label{sec:tests:results:validation}

While we demonstrated  that  we could reproduce the simulation results with a calibrated sub-grid model in \S~\ref{sec:tests:results:calibration}, a natural worry is that this calibration was simply fitted to one snapshot of one simulation, and cannot be extrapolated elsewhere -- essential if we wish to apply the sub-grid model to new simulations. We next therefore {\em validate} the sub-grid model in four different ways.

First, we compare snapshots at different times in our reference case from \S~\ref{sec:tests:results:calibration}, fixing $\kappa_{0}$ and $v_{\rm st,\,eff}$. Second, we compare simulations of other MW-mass galaxies (with the identical physics and numerical code, and same assumptions about fixed true physical $\bar{\nu}$ or parallel $\kappa_{\|}$ in code and $\bar{v}_{A}=v_{A}$), using the same coefficients. Third, we compare simulations of galaxies at different mass scales, in particular dwarf galaxies, where $v_{A}$ and field morphologies could be totally different (as the magnetic field strengths and gas densities and galaxy morphologies are different), again using the same code and same physical assumptions. If our sub-grid model is to be successfully applied for an entire cosmological simulation (across different mass and redshift scales as a galaxy forms), then it must be able to reproduce these situations reasonably well without ``re-calibration.'' Fig.~\ref{fig:compare.profiles} shows that it can indeed do so. Note we do not bother showing the comparison of different snapshot times for simulation {\bf m12i}, as the differences in time for that run are always smaller than the galaxy-to-galaxy differences.

Thus far these simulations all adopted the same true $\bar{\nu}$ (i.e.\ $\kappa_{\|}$) in-code. If our sub-grid model is robust, a fourth calibration is to consider models with systematically different $\kappa_{\|}$. For the reasons above (\S~\ref{sec:tests:results:calibration}), we expect some systematic offset between $\kappa_{0}$ and $\kappa_{\|}$. But if this is -- to lowest order -- a global systematic change between simulations,  then if we consider simulations with $10\times$ larger $\kappa_{\|}$, our inferred $\kappa_{0}$ should also increase by a factor of $\sim10$. Fig.~\ref{fig:compare.profiles} shows that this indeed works fairly accurately. 

Altogether, this suggests that we can approximate $P_{\rm cr}$ to within a factor of a few across a broad range of galaxy types, cosmological times, and plausible phenomenological (constant) diffusion/streaming coefficients, with a universal constant $\kappa_{0}$ and $v_{\rm st,\,eff}$ calibrated from full CR dynamics simulations.

\subsubsection{Application to Models with Non-Constant Diffusivity}
\label{sec:tests:results:variable}

In Fig.~\ref{fig:compare.profiles.variable} we explore whether our simple sub-grid model can be applied even to model situations where the diffusion coefficient (or equivalently CR scattering rate) is a complicated and highly-variable function of local plasma properties. We compare to three simulations from \citet{hopkins:cr.transport.constraints.from.galaxies,hopkins:2020.cr.transport.model.fx.galform} as noted above, which are constrained to be among the few in the set of models studied therein which are consistent with observational constrains on $\gamma$-ray emission and secondary-to-primary ratios in the Galaxy. In each of the three models plotted, the local scattering rates can vary by up to $\sim 10$ orders of magnitude in the ISM, and the functional dependence on e.g.\ strength of turbulence or magnetic fields is different in the different models. But qualitatively, in each of these models, the mean effective diffusivity rises as a function of galacto-centric radius on average in the CGM, to values up to $\gtrsim 10^{31}\,{\rm cm^{2}\,s^{-1}}$ at radii $\gtrsim 100\,$kpc -- much more consistent with the lower limits to the effective diffusivity in the distant CGM and IGM required by HST-COS observations around Milky Way-mass galaxies as shown in \citet{butsky:2022.cr.kappa.lower.limits.cgm}. For each, we treat the ``effective'' $\kappa_{0}$ and $v_{\rm st,\,eff}$ as parameters to be fit to the profiles (we show the values which give a best gas-mass-weighted fit to the profiles outside $>1\,$kpc), obtaining $(\kappa_{0}/{\rm cm^{2}\,s^{-1}},\,v_{\rm st,\,eff}/{\rm km\,s^{-1}}) =$ ($10^{28}$,150), ($10^{28}$, 100), ($2\times10^{30}$, 200) for runs ``\Alf-Max'', ``Fast-Max'', and ``SC:$f_{\rm QLT}=100$'', respectively. With this caveat, we see that all of the (observationally-allowed) models are sufficiently ``diffusive'' on average that they form a quasi-spherical profile that can be reasonably approximated by our sub-grid model for a given galaxy and time. The major caveat is that because the coefficients depend systematically on certain properties, there is no guarantee this would extrapolate to different galaxy masses or redshifts (or regions of the ISM with extreme parameters), without some physical prior (for example knowing the dependence on some local property like magnetic field strength, and then folding in some assumption on how the magnetic field strength depends on galaxy mass and redshift). Indeed, as shown in \citet{hopkins:cr.transport.constraints.from.galaxies}, some of these models produce systematically different ``effective diffusivity'' in small dwarf galaxies, despite this number being broadly similar in the ISM of the Milky Way.

\subsubsection{Observational Constraints on Model Choices}
\label{sec:tests:results:obs.input}

Of course, one cannot simply adopt totally-arbitrary values of $\kappa_{0}$ or $v_{\rm st,\,eff}$ within our model and remain consistent with observations. As discussed extensively in \citet{hopkins:cr.transport.constraints.from.galaxies}, only a small subset of the possible model space of CR transport parameters in the ISM of Milky Way-like, low-redshift galaxies are consistent with the combination of observations constraining $\sim1-10\,$GV protons, including their observed spectra and energy densities, primary-to-secondary and radioactive isotope ratios (constraining their grammage and residence times) around the Solar neighborhood, as well as $\gamma$-ray emission constraints integrated over our Galaxy as well as other Local Group galaxies. The ``viable'' models from detailed comparison to these constraints include those in Fig.~\ref{fig:compare.profiles} with $\kappa_{\|} \sim 3\times10^{29}\,{\rm cm^{2}\,s^{-1}}$, as well as the models in Fig.~\ref{fig:compare.profiles.variable}, which together fall into a relatively narrow range of allowed ``effective isotropically-averaged diffusivity around the Solar circle,'' $\kappa_{\rm iso,\,eff}^{\odot} \sim \kappa_{0} + v_{\rm st,\,eff}\,r_{\odot}^{\rm gal}$.

It is not our intention here to derive or re-make these arguments, but we do wish to emphasize that within the context of our simple model where $\kappa_{0}$ and $v_{\rm st,\,eff}$ are global constants, observations actually do constrain their approximate sum relatively well, requiring that the {\em isotropically-averaged} sum obeys:
\begin{align}
\label{eqn:kappa.constraint} \kappa_{\rm iso,\,eff}^{\odot} \equiv \kappa_{0} + v_{\rm st,\,eff}\,r_{\odot}^{\rm gal} \sim 10^{29}\,{\rm cm^{2}\,s^{-1}}
\end{align}
with $r_{\odot}^{\rm gal}  \sim 8\,$kpc. Again we stress that this is remarkably consistent with the constraint from many other recent studies of CR propagation in the Milky Way using classic models like GALPROP to make detailed models of the entire CR spectrum of all species observed, such as e.g.\ \citet{evoli:2019.cr.fitting.galprop.update.ams02} or \citet{korsmeier:2022.cr.fitting.update.ams02}, provided one properly accounts for the integration over all CR energies, difference between isotropic and anisotropic coefficients, and includes all of the relevant streaming/advective/convective/diffusive velocity terms sometimes separated in those models.

So ensuring that Eq.~\ref{eqn:kappa.constraint} is at least approximately (order-of-magnitude) satisfied in the ISM helps to ``anchor'' the simple sub-grid models proposed here to much more detailed simulation literature models (e.g.\ \citealt{chan:2018.cosmicray.fire.gammaray,hopkins:2020.cr.transport.model.fx.galform,buck:2020.cosmic.ray.low.coeff.high.Egamma,wehahn:2021.gamma.rays}) which attempt to simultaneously fit a wide range of more detailed observational constraints, including the CR energy density at the Solar circle observed by e.g.\ \citep{cummings:2016.voyager.1.cr.spectra} and $\gamma$-ray emission (both very crudely inversely proportional to the CR transport speed, for all else equal) observed in the Galaxy and nearby systems \citep{lacki:2011.cosmic.ray.sub.calorimetric,tang:2014.ngc.2146.proton.calorimeter,griffin:2016.arp220.detection.gammarays,fu:2017.m33.revised.cr.upper.limit,wjac:2017.4945.gamma.rays,wang:2018.starbursts.are.proton.calorimeters,lopez:2018.smc.below.calorimetric.crs}, as well as even more detailed fits to full CR spectra in the Solar circle including their residence times, radioactive isotope and secondary-to primary ratios \citep[e.g.][]{blasi:cr.propagation.constraints,vladimirov:cr.highegy.diff,gaggero:2015.cr.diffusion.coefficient,2016ApJ...819...54G,2016ApJ...824...16J,cummings:2016.voyager.1.cr.spectra,2016PhRvD..94l3019K,evoli:dragon2.cr.prop,2018AdSpR..62.2731A,hopkins:cr.multibin.mw.comparison,delaTorre:2021.dragon2.methods.new.model.comparison}.

That said, there is still some degeneracy (at the extremely simple level of the model here) between $\kappa_{0}$ and $v_{\rm st,\,eff}$ in the ISM allowed by different models which fit present observations comparably well: for example Eq.~\ref{eqn:kappa.constraint} can be satisfied by a model with $\kappa_{0} \sim 10^{29}\,{\rm cm^{2}\,s^{-1}}$ and small $v_{\rm st,\,eff}$, or a model with large $v_{\rm st,\,eff} \sim 40\,{\rm km\,s^{-1}}$ and small $\kappa_{0}$ (the equivalent of models with an effective constant parallel/anisotropic diffusivity at $\sim$\,GeV energies of $\kappa_{\|} \sim 3\times10^{29}\,{\rm cm^{2}\,s^{-1}}$ and  small $v_{\rm st,\,\|}$, or models with $v_{\rm st,\,\|} \sim 120\,{\rm km\,s^{-1}}$ -- i.e.\ highly ``super-\Alf{ic} streaming -- and small $\kappa_{\|}$). However, these observational constraints effectively reduce the ``degrees of freedom'' of these (simplified) sub-grid models to a single parameter, the relative ratio of $\kappa_{0}$ to $v_{\rm st,\,eff}$ in Eq.~\ref{eqn:kappa.constraint}. Other observational probes which have recently been proposed to constrain the effective run of $\kappa_{\rm iso,\,eff}$ with galacto-centric radius $r^{\rm gal}$ in the CGM, e.g.\ those in \citet{butsky:2022.cr.kappa.lower.limits.cgm}, can be used to further constrain this (which, for the simple models here, favors relatively large $v_{\rm st,\,eff} \sim 100-200\,{\rm km\,s^{-1}}$ so that $\kappa_{\rm iso,\,eff}$ rises significantly with galacto-centric radius).\footnote{One can also modify the assumed functional form of the CR ``effective diffusivity'' if desired: consider e.g.\ a model with $\langle \kappa_{\rm iso} \rangle \sim \kappa_{0} + a\,r^{m}$. We then simply replace $\kappa_{0} + v_{\rm st,\,eff}\,r_{ij} \rightarrow \kappa_{0} + a\,r_{ij}^{m}$ in Eq.~\ref{eqn:cr.egy.touse}. For $m<2$ we then take $\psi_{\rm loss} \rightarrow (\Psi_{\rm loss}/a)^{1/(2-m)}\,[1 + \kappa_{0}\,\Psi_{\rm loss}^{m/(2-m)}\,a^{-2/(2-m)} ]^{-1/2}$ in Eq.~\ref{eqn:psi.loss}, and $r_{\rm max} \rightarrow (a\,t_{\rm max}/2)^{1/(2-m)}\,[1+\{1 + (16\,\kappa_{0}\,t_{\rm max}/[a\,t_{\rm max}]^{2/(2-m)}) \}^{1/2}]$ in Eq.~\ref{eqn:rmax}. For $m\ge 2$ the solutions take a different form and we can approximate $\Delta \tau_{{\rm cr},\,i} \rightarrow (1/2)\,{\rm MIN}[ (\Psi_{\rm loss}/\kappa_{0})^{1/2}\,\ell \ , (\Psi_{\rm loss}/\kappa_{0})\,(\kappa_{0}/a)^{2/m} ]$ (with $\ell \equiv (\Delta x_{i}^{2} + (\rho_{{\rm gas}\,i}/|\nabla \rho_{{\rm gas},\,i}|)^{2})^{1/2}$)  in Eq.~\ref{eqn:psi.loss}, and $r_{\rm max} \rightarrow 2\,(\kappa_{0}\,t_{\rm max})^{1/2}$ for $t_{\rm max} \le \kappa_{0}^{2/m-1}\,a^{-2/m}$ or $r_{\rm max} \rightarrow \infty$ otherwise, in Eq.~\ref{eqn:rmax}. With this formulation, we can better fit the constraints in  \citet{butsky:2022.cr.kappa.lower.limits.cgm} by adopting $m=2$, $\kappa_{0} \sim 5\times10^{28}\,{\rm cm^{2}\,s^{-1}}$ and $a \sim 2\,{\rm Gyr}^{-1}$.} It may also be possible to constrain these with more detailed models of Solar neighborhood CRs incorporating the full CR spectrum and diffuse galactic emission, though if $\kappa$ or $v_{\rm st,\,eff}$ is spatially-variable, or time-dependent terms in the transport equations are important, or the effective CR scattering halo properties vary with rigidity, such separation could prove more challenging (see discussion in \citealt{hopkins:cr.multibin.mw.comparison}).

\subsubsection{Limitations \&\ Failures}
\label{sec:tests:results:failures}

Despite its surprising success the simplified toy model here has many limitations. Obviously the model here cannot and should not be used to predict detailed CR observables (e.g.\ $\gamma$-ray or synchrotron emission, let alone detailed primary-to-secondary or isotopic ratios) which depend sensitively on spectrally-dependent transport physics and terms that we have explicitly neglected here \citep[see][]{chan:2018.cosmicray.fire.gammaray,hopkins:cr.multibin.mw.comparison,hopkins:2021.sc.et.models.incompatible.obs}. 

At very low transport speeds ($\kappa_{0} \ll 10^{29}\,{\rm cm\,s^{-1}}$), our sub-grid model approximations break down, as the CRs become more tightly-coupled to the gas, so terms we have neglected such as adiabatic gains/losses, advection with gas, anisotropic and  globally non-spherical behaviors (e.g.\ being trapped with the gas in a thin disk instead of diffusing to a thicker distribution), and out-of-equilibrium behaviors (whose convergence time is longer at lower $\kappa$) become progressively more important. Fortunately, this appears to be clearly ruled-out by present CR observations (\S~\ref{sec:tests:results:obs.input}), for the $\sim$\,GeV CRs which dominate the total CR energy/pressure -- i.e.\ our model assumptions work because observations appear to require reasonably high CR diffusivities. However, it could still be the case that very low-energy CRs ($\lesssim 10\,$MeV), which are not important for the total CR pressure or energy density but dominate the CR ionization rate, could be much more tightly-coupled to the gas (see e.g.\ \citealt{indriolo:2009.high.cr.ionization.rate.clouds.alt.source.models,padovani:2009.cr.ionization.gmc.rates.model.w.alt.sources,indriolo:2012.cr.ionization.rate.vs.cloud.column,hopkins:cr.multibin.mw.comparison}). This means that care is needed assuming that one can extrapolate from this sub-grid model (or any ``single-bin'' CR model) to much lower-energy CR dynamics. And of course if real CR transport parameters depend on local plasma properties there could be regimes of parameter space where the diffusivity is much lower.

Similarly, our sub-grid model performs relatively poorly when CRs approach the proton calorimetric limit  -- i.e.\ when most of the CR energy is lost before CRs escape dense gas in their galactic vicinity. For very low diffusivities as above, this would occur in all galaxies (including Milky Way like through even small dwarf galaxies), but as noted this is not observed. However even for the observationally-favored diffusivities, this can and is observed to occur in extremely dense starburst galaxies \citep{lacki:2011.cosmic.ray.sub.calorimetric,tang:2014.ngc.2146.proton.calorimeter,griffin:2016.arp220.detection.gammarays,fu:2017.m33.revised.cr.upper.limit,wjac:2017.4945.gamma.rays,wang:2018.starbursts.are.proton.calorimeters,lopez:2018.smc.below.calorimetric.crs}. In this limit our simple treatment of CR attenuation via local and self ``shielding'' is not particularly accurate (similar to the weakness in the LEBRON method for RHD, discussed in \citealt{hopkins:radiation.methods}). However, essentially by definition in such a case, CRs pressures are exponentially suppressed so we are safely in the limit where CR pressure is a tiny portion of total pressure -- hence getting the CR pressure exactly ``correct'' in such a case is not particularly important for predicting CR effects on galaxies. But it is important to attempt to capture the ``transition'' to this regime -- without some accounting for CR  losses, one could severely over-estimate the importance of CR pressure in dense starburst or high-redshift systems. Our simulation calibration sample largely does not reach these extreme conditions, so future tests in this regime would be particularly useful.

Additionally, as discussed in \S~\ref{sec:tests:results:calibration}, our sub-grid model also becomes notably less accurate and robust ``inside'' the galaxy, specifically at locations ``between'' a number of comparable-strength sources (e.g.\ modeling variations in CR properties within the galaxy ISM at sub-kpc scales). This is not surprising, as assumptions like spherical symmetry are much more accurate on larger scales, and our model is really intended to accurately capture the behavior of CR pressure on scales ``outside'' a collection of dominant sources -- e.g.\ in the CGM around a galaxy. This is an important caveat for a wide range of possible applications.

\subsection{Application in a Full Cosmological Simulation}
\label{sec:tests:results:application}

We now test an actual ``on-the-fly'' application of the sub-grid model. Specifically, we re-run the simulation {\bf m11i} with $\kappa_{\|}=3\times10^{29}\,{\rm cm^{2}\,s^{-1}}$ analyzed in Fig.~\ref{fig:compare.profiles}, from its cosmological initial conditions at $z=100$, turning off the full on-the-fly dynamical explicit evolution of the CR energy density and flux used in the run in Fig.~\ref{fig:compare.profiles}, and instead using our proposed sub-grid model from \S~\ref{sec:numerical} on the fly. We use the sub-grid model at every timestep and cell to compute $e_{\rm cr}({\bf x},\,t)$, then use this CR energy density throughout the code to compute CR pressure forces, heating, and ionization rates identical to what is done in the ``full CR dynamics'' simulation (so the CR heating, pressure effects, etc., are identical {\em for a given} $e_{\rm cr}({\bf x},\,t)$ in the two runs, but we have replaced the actual CR transport algorithm). For reference, we also compare to the ``pure hydrodynamics'' version of the same simulation, which was studied and compared in detail to the full CR dynamics run in \citet{hopkins:cr.mhd.fire2}. That run adopts identical physics and  numerics, but simply disables the explicit CR dynamics and MHD entirely. In Fig.~\ref{fig:cosmo.test}, we compare some basic properties (a subset of those studied in detail in \citealt{hopkins:cr.mhd.fire2}) in the simulation using the sub-grid model and full CR dynamics. Most importantly, \citet{hopkins:cr.mhd.fire2} show that in the full CR run (compared to ``no CRs'' runs with or without magnetic fields), the additional pressure provided by CRs in the CGM suppresses accretion onto the galaxy, in turn suppressing the star formation rate and stellar mass of the galaxy by redshift $z=0$ (we chose galaxy {\bf m11i} in part because it exhibits some of the most dramatic effects of CRs of any galaxy simulated in \citealt{hopkins:cr.mhd.fire2}). This is the most important aspect of CRs for our sub-grid model to capture. 

The sub-grid model appears to capture the leading-order effects of CRs on galaxy formation remarkably well, at least in this particular case. The star formation rate (SFR) and stellar mass growth history are very similar to the ``full CR dynamics'' run (modulo stochastic effects such as the detailed amplitude and timing of individual ``bursty'' star formation events). The stellar mass and late-time SFR in the subgrid run is slightly lower than the ``full CRs'' run, but this offset is modest and results in a very small ($\sim0.1\,$dex) stellar mass difference at $z=0$, within the range of run-to-run stochastic variations \citep{su:discrete.imf.fx.fire,keller:stochastic.gal.form.fx,genel:stochastic.gal.form.fx}. Following the star formation history, the metallicity and rotation curve and baryonic mass distribution (stellar effective radius and baryonic mass profile) are also reasonably well-reproduced. The subgrid CR run does predict a lower metallicity compared to the explicit CR run, consistent with the slightly lower late-time SFR in the subgrid run,  suggesting that the sub-grid CR pressure in the CGM (probably somewhat over-estimated) may have prevented some late-time recycling of enriched material which would otherwise have been re-accreted and formed stars later (boosting their metallicity) in the full/explicit CR dynamics run. Unfortunately the difference is much smaller than typical systematic uncertainties in both metallicity measurements \citep{tremonti:mass.metallicity.relation} and metal yield models or other galaxy formation ``input'' physics \citep{ma:2015.fire.mass.metallicity}. Another related discrepancy between ``full CR dynamics'' and sub-grid models in Fig.~\ref{fig:cosmo.test} appears to be the weaker ``core'' in the central dark matter density  profile at $<1\,$kpc in the subgrid model. This again is consistent with the stellar  mass and SF history differences, as at precisely these masses the core strength is a strong function of stellar mass -- more late star formation generates a stronger core \citep[see][]{onorbe:2015.fire.cores,chan:fire.dwarf.cusps,lazar:cored.dm.mass.profiles}. Interestingly, a stronger core {\em is} present earlier in the subgrid-model run (when its SFR is closer to the explicit CR dynamics run), which then slightly ``fills in'' as the late-time star formation ``bursts'' are too weak to re-excavate it (similar to the process described in \citealt{onorbe:2015.fire.cores}). 

All of these suggest that the sub-grid model performs reasonably well, though it may slightly over-estimate the effects of CR feedback at late cosmic times (perhaps in part because the CRs are not infinitely diffusive and therefore not perfectly smoothly-distributed in the CGM, and their inhomogeneity modifies the thermal instability of that medium in turn; see \citealt{butsky:2020.cr.fx.thermal.instab.cgm}). And in particular this reinforces the idea that the subgrid model should not be taken too seriously at significantly sub-kpc scales. 

But in terms of computational cost: the ``full CRs'' model is nearly an order of magnitude more computationally expensive than the ``no CRs'' run, primarily owing to timestep limitations as discussed in \S~\ref{sec:intro}. Meanwhile, the ``subgrid'' model is actually {\em faster} than the ``no CRs'' run by a factor of $\sim 2-3$, owing to the reduced stellar mass and star formation rate (less dense gas and fewer SNe, which require small timesteps). Thus while imperfect, the subgrid model is able to capture some of the dominant CR effects at radically reduced computational expense, as desired.

\section{Conclusions}
\label{sec:discussion}

We have presented an intentionally extremely-simplified sub-grid model for CR ``feedback'' in galaxy formation simulations or semi-analytic models, designed to impose negligible computational cost. We derive the model beginning from exact expressions for CR transport and dynamics, then making successive simplifying assumptions until we arrive at a simple, equilibrium steady-state isotropic analytic expression \S~\ref{sec:deriv}. We then present a detailed numerical implementation which can be solved trivially alongside gravity in most numerical codes capable of galaxy-scale simulations or even purely-analytically in semi-analytic models in spherical symmetry (\S~\ref{sec:numerical}). Given the many simplifying assumptions needed to render the model tractable and low-cost, we stress that caution is needed, and we discuss the relevant approximations and assumptions systematically in order to note where they will break down and where the dominant uncertainties in using this model will arise (\S~\ref{sec:assumptions}). We then validate the model by comparing to detailed high-resolution simulations which explicitly follow anisotropic diffusion and streaming of CRs from individual SNe and shocked stellar winds in CR-MHD galaxy formation simulations (\S~\ref{sec:tests}) to show that the most important effects of CRs can be reasonably captured to at least order-of-magnitude accuracy with this toy model. 

We stress that our goal here is to enable at least {\em some} exploration of the effect of CRs in simulations/semi-analytic models which either do not model CR transport explicitly owing to numerical limitations or cannot do so realistically owing to limited resolution, lack of magnetic fields, lack of explicit treatment of neutral ISM/CGM phases, or other challenges. We attempt to derive the simplest-possible model that captures the leading-order effects of CRs on galaxy dynamics -- the models here are clearly not suitable for predicting CR observables ($\gamma$-ray or synchrotron emission, secondary-to-primary ratios, etc.), nor for capturing more subtle effects (CR energy dependence and small-scale ISM variations in ionization rates, effects of CRs within acceleration regions and individual SNe bubbles, etc.). And the sub-grid model is designed to be accurate on relatively large coarse-graining scales in both space ($\gtrsim$\,kpc) and time ($\gtrsim\,$Myr), appropriate for galactic astrophysics and cosmology. In future work, it will be interesting to explore the effects of the CR model here in various contexts with broad parameter surveys, coupled to a smaller number of explicit CR-MHD simulations to follow the dynamics in more detail and better-calibrate the models. And future work understanding the CR transport coefficients themselves is still needed to inform all CR models. Here we parameterize our model with constant transport coefficients, which can reproduce many CR observations if the effective isotropic diffusivity is chosen within a narrow range around $\kappa_{\rm iso,\,eff}^{\odot} \sim 10^{29}\,{\rm cm^{2}\,s^{-1}}$, but we stress that the approximation of constant in space-and-time CR transport coefficients is almost certainly incorrect in detail. But the true scaling of CR transport properties with local plasma properties remains deeply uncertain, and is a subject beyond our study here.

\acknowledgments{Support for PFH was provided by NSF Research Grants 1911233, 20009234, 2108318, NSF CAREER grant 1455342, NASA grants 80NSSC18K0562, HST-AR-15800. Numerical calculations were run on the Caltech compute cluster ``Wheeler,'' allocations AST21010 and AST20016 supported by the NSF and TACC, and NASA HEC SMD-16-7592.}

\datastatement{The data supporting this article are available on reasonable request to the corresponding author.} 

\bibliography{ms_extracted}
\end{document}